\documentclass[twocolumn,aps,prd,showpacs,groupedaddress]{revtex4}

\usepackage{amsmath}
\usepackage{dcolumn}
\usepackage{epsfig}
\usepackage{graphicx}
\usepackage{latexsym}

\usepackage{epstopdf}

\begin{document}
\hyphenation{Ka-pi-tul-nik}

\title{A New Apparatus for Detecting Micron-Scale Deviations from Newtonian Gravity}

\author{David M. Weld$^1$}
\email{dweld@mit.edu}
\thanks{Present address: Center for Ultracold Atoms, Massachusetts Institute of Technology, Cambridge MA 02139}
\author{Jing Xia$^1$}
\author{Blas Cabrera$^1$}
\author{Aharon Kapitulnik$^{1,2}$}
\affiliation{$^1$Department of Physics, Stanford University, Stanford, CA 94305}
\affiliation{$^2$Department of Applied Physics, Stanford University, Stanford, CA 94305} 

\begin{abstract}

We describe the design and construction of a new apparatus for detecting or constraining deviations from Newtonian gravity at short length scales.  The apparatus consists of a new type of probe with rotary mass actuation and cantilever-based force detection which is used to directly measure the force between two micromachined masses separated by tens of microns.  We present the first data from the experiment, and discuss the prospects of more precisely constraining or detecting non-Newtonian effects using this probe.  Currently, the sensitivity to attractive mass-dependent forces is equal to the best existing limits at length scales near 5 $\mu$m.  No non-Newtonian effects are detected at that level.


\end{abstract}

\pacs{04.80.Cc, 04.50.-h, 07.10.Cm} 

\maketitle

\section{INTRODUCTION}

\subsection{Motivation}

In spite of the success of the Newtonian and Einsteinian theories of the gravitational interaction, gravity remains the most mysterious of the four fundamental forces.  No renormalizable quantum field theory of gravity has been found, and gravity plays a central role in two of the greatest outstanding questions in physics: the cosmological constant problem~\cite{beane,sundrumCC} and the gauge hierarchy problem~\cite{adelbergerinvsq}.  It is generally believed that a classical description of gravity must fail at or before the Planck length $\sqrt{\hbar G/c^3}$  (about $10^{-35}$ meters).  However, many of the proposed solutions to these problems predict modifications to Newtonian gravity at much longer length scales, up to 1 mm.  A few of the most important theoretical proposals along these lines are  ``large'' extra spatial dimensions~\cite{add1,add2, antoniadis1, antoniadis2}, string theoretic  dilatons and moduli~\cite{dilaton}, and scalar particles in ``hidden'' supersymmetric sectors~\cite{adelbergerinvsq}.  Additionally, it has been pointed out~\cite{sundrumCC}  that the observed value of the cosmological constant, when expressed as a length scale $(\hbar c / \Lambda)^{1/4}$, is about 100 $\mu$m.  There are many reasons to investigate the behavior of gravity at small length scales, perhaps the most compelling of which is the fact that it is largely unexplored territory.

Until recently, Newtonian gravity had only been experimentally tested down to millimeter length scales; the extreme weakness of gravity that poses such a puzzle for theorists is an equally daunting problem for experimentalists.  In recent years, however, there have been several experiments designed to detect or constrain deviations from Newtonian gravity below 1 millimeter~\cite{frogland1,adelbergerexpt1,priceexpt,fischbachexpt,lamoreauxalphlam,newadelberger}.  Typically, the results of such experiments have been reported as bounds on a hypothetical Yukawa coupling added to the Newtonian potential.  This modified potential takes the form
\begin{equation}
V=-\frac{G m_1 m_2}{r} \left( 1 + \alpha \, e^{-r/\lambda}\right),
\label{yukawaforce}\end{equation}
where $\alpha$ represents the strength (relative to gravity) of the Yukawa coupling, and $\lambda$ represents its length scale.  Such Yukawa couplings arise naturally from interactions due to exotic massive scalar particles; tests of gravity at or near the length scale of any extra dimensions would also see corrections to Newtonian gravity of this form~\cite{add2}.  Any given experiment of this type has a geometry-dependent range of $\lambda$ to which it is most sensitive.  The finite force sensitivity of any measurement will (if no non-Newtonian effects are detected) result in an upper bound for the strength of any such effect, which can be expressed as an upper bound for $\alpha$ as a function of $\lambda$.    Thus, each experiment carves out an area of \mbox{$\alpha$-$\lambda$} space within which it would be able to detect deviations from Newtonian gravity.  

\subsection{Overview}

\begin{figure}[tbh!]\begin{center}
\includegraphics[width=\columnwidth]{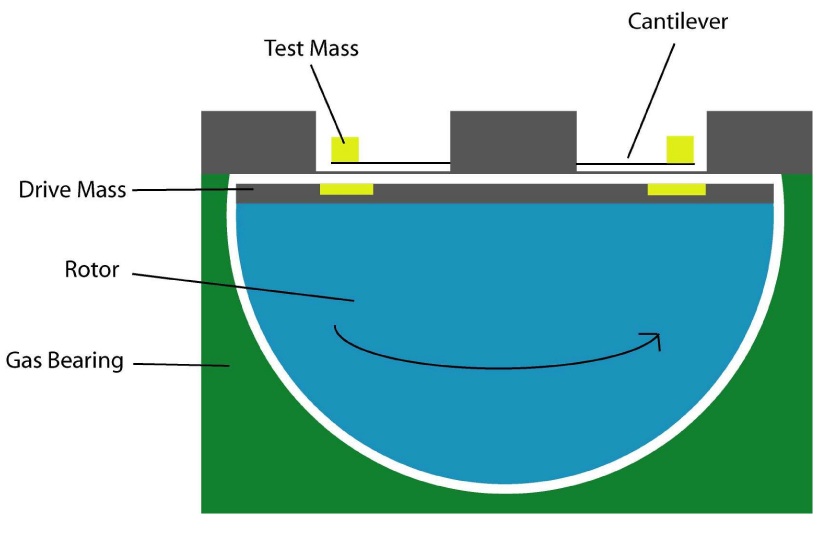}
\caption[Schematic of the experiment]{(color online). A schematic side view (not to scale) of the heart of the experimental apparatus.}
\label{schematic1}
\end{center}\end{figure}

\begin{figure}[tbh!]\begin{center}
\includegraphics[width=0.75 \columnwidth]{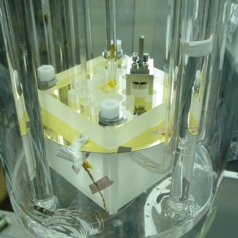}
\caption[Assembled probe]{(color online). A photograph of the assembled probe.  The gold-coated cantilever wafer is visible between the lid and housing of the gas bearing.  The stainless steel structure protruding from the lid is the fiber holder that maintains the alignment of the interferometer.  The optical fiber is barely visible emerging from the top of that structure.  The fiber bundle for the encoder and two of the gas bearing inlets are also visible here.}
\label{assembledprobe}
\end{center}\end{figure}

Fig.\ \ref{schematic1} presents a schematic, and Fig.\ \ref{assembledprobe} a photograph, of the new experimental apparatus.  It is fundamentally a Cavendish-type~\cite{cavendish} experiment in the sense that its purpose is to directly measure the force between two masses.  A cryogenic helium gas bearing \cite{felch} is used to rotate a disc containing a drive mass pattern of alternating density under a small test mass mounted on a micromachined cantilever.  Any mass-dependent force between the two will produce a time-varying force on the test mass, and consequently a time-varying displacement of the cantilever.  This displacement is read out with a laser interferometer, and the position of the drive mass is simultaneously recorded using an optical encoder.  The displacement is then averaged over many cycles and converted to a force using measured properties of the cantilever.  This AC ``lock-in'' type measurement enables significant noise rejection and allows us to operate on resonance to take advantage of the cantilever's high quality factor.   
\subsection{Organization of the paper}

Sec.~II of this paper discusses the experimental apparatus in detail.  Sec.~III elucidates the data acquisition and averaging procedures used to extract the measured force limits.  Sec.~IV presents the experimental uncertainties and discusses noise and background force considerations.  Sec.~V contains analysis of the first results obtained with the new apparatus, and conclusions and future prospects are discussed in Sec.~VI.

\section{EXPERIMENTAL APPARATUS \label{apparatuschapter}}

Conceptually, the experiment can be divided into four parts: the drive mass actuation system, the drive mass, the test mass, and the cantilever.  Each of those subsystems is discussed below.

\subsection{Drive Mass Actuation}

\begin{figure}[tbh!]\begin{center}
\includegraphics[width=0.45 \columnwidth]{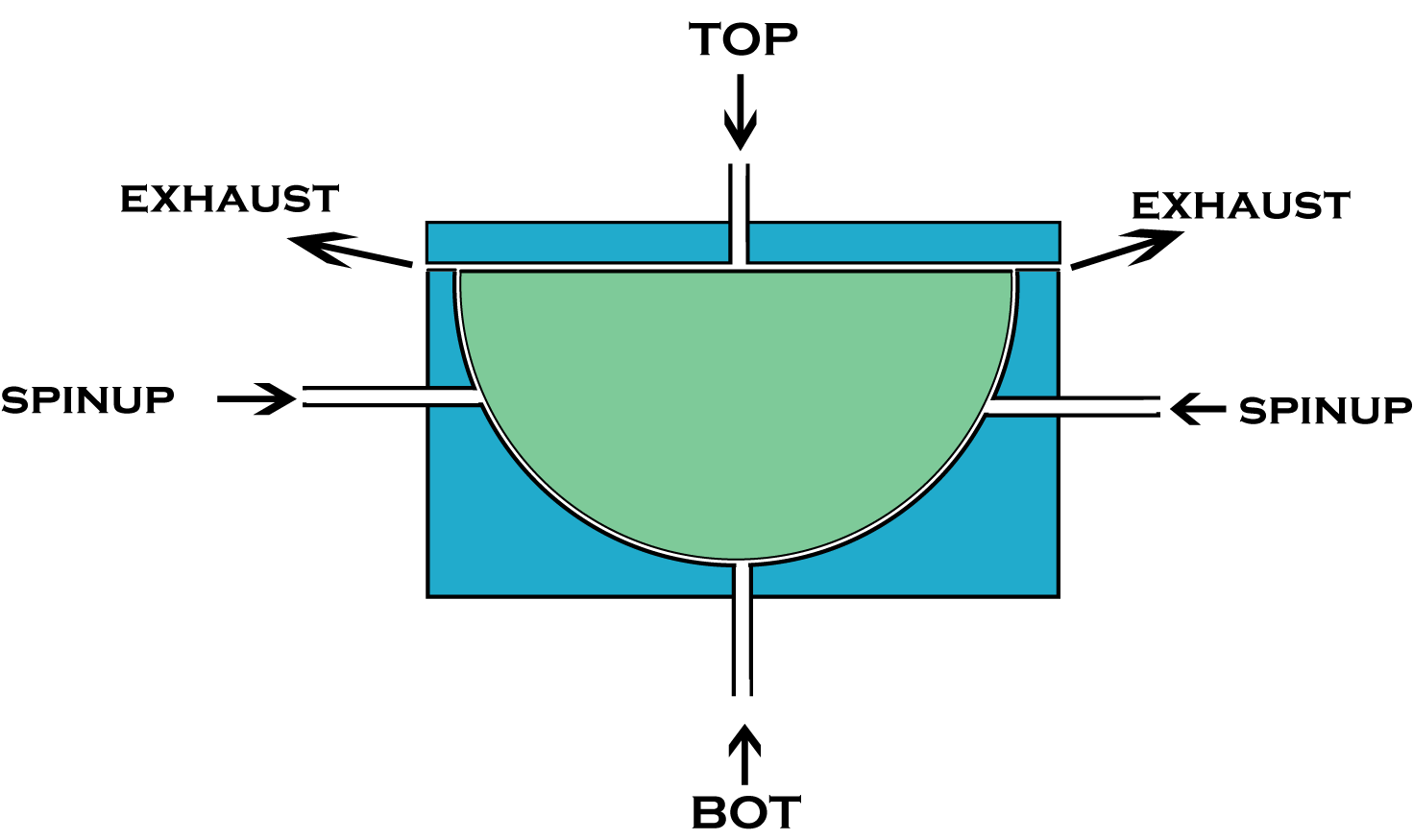} 
\includegraphics[width=0.45 \columnwidth]{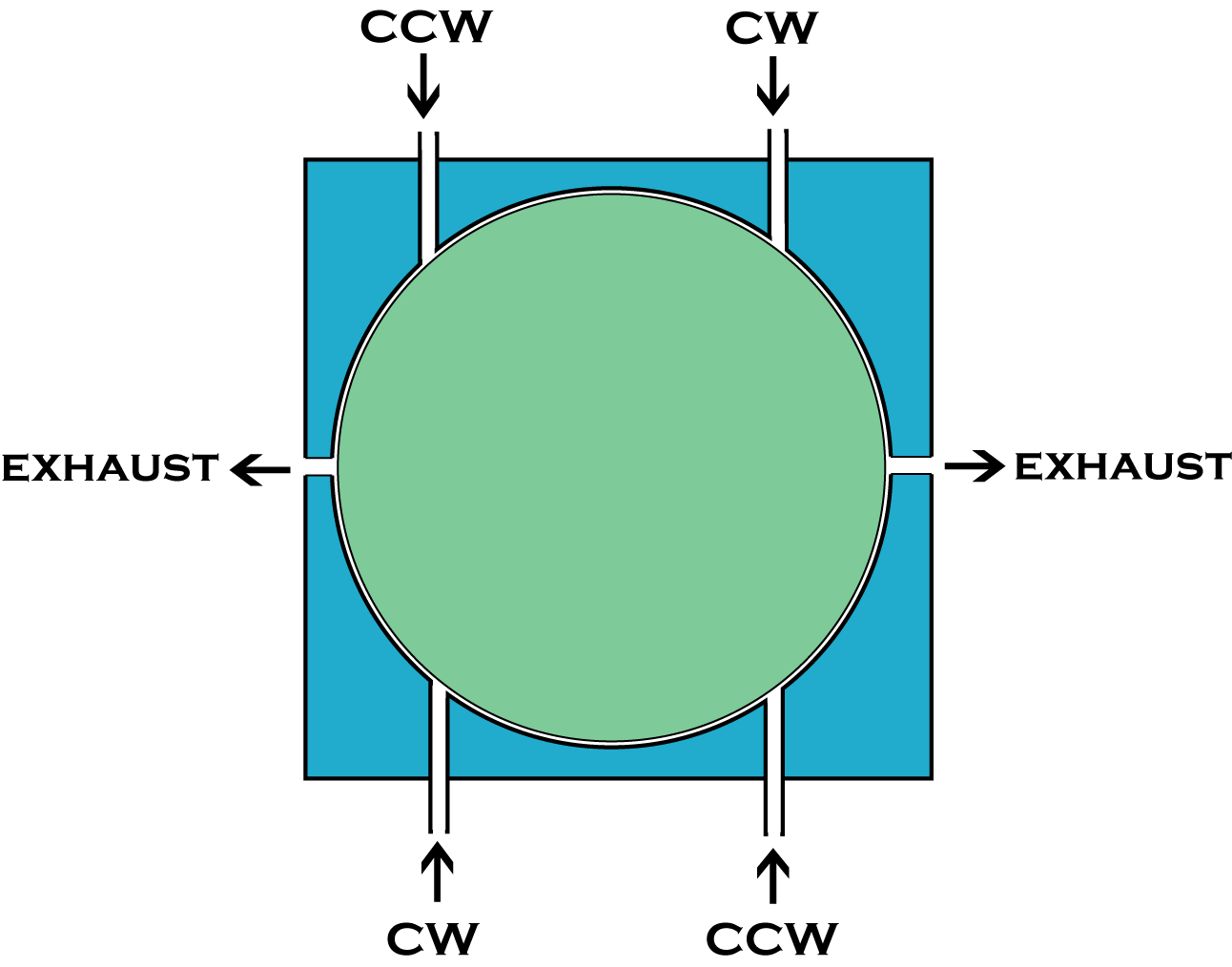}
\caption[Schematic of the gas bearing]{(color online). A schematic drawing of the helium gas bearing, showing the six gas inlet ports.  \textbf{Left:} Side view.  \textbf{Right:} Top view.  The ports labeled ``TOP'' and ``BOT'' provide the gas that the rotor floats on.  The clockwise and counter-clockwise spin-up ports are labeled ``CW'' and ``CCW'' respectively.  Exhaust ports are also indicated.}
\label{gasbearing}
\end{center}\end{figure}

The mass actuation system (schematically depicted in Fig.\ \ref{gasbearing}) consists of a rotary helium gas bearing, which is a two-inch-diameter hemispherical cavity with six gas inlets---two for the bearing and four for spin control.  A hemispherical rotor sits within the cavity; it can be levitated and continuously rotated by adjusting the flow of gas among the inlets.  Computer-interfaced mass flow controllers are combined with the spin-speed detection system into a feedback loop to keep the rotor spinning at a set frequency.  At low temperature, the gas flow is laminar in all of the bearing except for the spin-up channels and the exhaust \cite{felch}.  The wafer containing the cantilevers is clamped under the flat of the hemispherical cavity, and the drive mass is mounted on the flat of the rotor.  Both the cavity and the rotor are machined from single blocks of fused silica (used for its very low thermal expansion).  This type of motor/bearing assembly has many advantages for this experiment; we discuss some of them below.  
\subsubsection{Advantages of the Gas Bearing}

The cryogenic fused quartz helium gas bearing used in this experiment was originally built and used to measure the Cooper pair mass in niobium \cite{felch, tate1, tate2}.  That experiment measured the zeros of the total magnetic flux (London moment plus or minus integer numbers of flux quanta) through a superconducting niobium ring deposited around the equator of the slightly-more-than-hemispherical rotor.  The extremely precise and stable control and readout of the rotor's rotation frequency that was achieved in that work is also useful for the present experiment.

Two important virtues of the bearing are its tight distance tolerances and its stiffness; run in the correct regime, it can be substantially stiffer than a bearing made of steel.  In an experiment designed to detect mass-dependent forces (particularly exponentially decaying Yukawa forces) it is critically important to know and control exactly how far apart the test mass and drive mass are.  In the configuration of our experiment, that corresponds to knowing and controlling the separation between the top flat surface of the rotor and the bottom surface of the bearing's lid (which in our case is a silicon cantilever wafer).  The gas bearing performs very well in this regard, in that if the gas flows are correctly adjusted it strongly resists any attempts to compress the gas layer between rotor and lid below about 12 $\mu$m in thickness.  This is due to the fact that pressure is inversely proportional to the thickness of this gas layer, assuming a constant mass flow rate.  The restoring force thus diverges as the separation gets very small, which results in a high stiffness in the relevant direction.  This stiffness means that not only is it relatively immune to noise-producing ``rattling'' displacements, it also is capable of robust and repeatable operation in a distance regime that is very interesting from the point of view of theoretical predictions of deviations from Newtonian gravity.

Other advantages of using the cryogenic helium gas bearing as a drive mass actuator stem from what it is not.  It is totally non-magnetic, a characteristic which distinguishes it from most low-temperature electric motors.  Since the command signals that control drive mass motion are sent in the form of slowly-varying gas flows rather than AC voltages, it is also completely immune to in-probe electrical cross-talk.  Because of its unique system of gaseous lubrication and actuation, the bearing can be easily operated at any temperature above the boiling point of liquid helium where the viscosity of helium gas is very low; other types of bearings tend to run into serious lubrication problems at very low temperatures.

The rotary geometry of the gas bearing is also important for our experiment.  By far the most important benefit of rotary actuation is that its large range of motion (compared to linear piezoelectric actuators) allows us to use very large area masses in a parallel-plate configuration.  Since the force due to any mass-dependent interaction scales with the area of the masses, this apparatus has an excellent intrinsic sensitivity to such interactions.  

Finally, the self-aligning nature of the gas bearing greatly simplifies the experimental procedure.  The tight clearances and stiffness of the bearing (and, secondarily, the large extent of the drive mass trenches in the radial direction) make it auto-aligning--- if the bearing is spinning, the the drive masses are guaranteed to be directly below the test masses.  This auto-alignment is more than an experimental convenience; it enables long averaging times and thus improved force sensitivity.

\subsubsection{Disadvantages of the Gas Bearing}

No other practically achievable drive mass actuation method has all the advantages discussed above.  However, the unique nature of the cryogenic gas bearing did pose some important experimental challenges not posed by other actuation methods.  The most important of these was the fact that it requires gas to operate, while the cantilevers require vacuum to maintain a high quality factor and thus a good force sensitivity.  This problem was solved by hermetically sealing the cantilevers inside a cryopumped cavity.

Another potential downside of the gas bearing is that it can produce substantial vibrational noise.  This noise is important because of the limitations on possible vibrational isolation imposed by our experimental geometry.  Indeed, random vibrational noise (and not thermal noise) is currently the limiting factor that determines the performance of the experiment.  However, the portion of this noise that is due to the gas flow in the bearing can be drastically reduced by reducing the gas flows in the top and bottom bearing inlets.  This reduction is presumably due to reduced turbulence at low gas flow rates.  This disadvantage, then, appears to be surmountable.
\subsubsection{Flow meters and flow controllers}
As depicted in Fig.\ \ref{gasbearing}, there are four independent gas lines running into the probe: one each for the top and bottom bearing inlets, and two others that are each split and fed to the two clockwise and two counter-clockwise spin-up inlets.  The flow rate of the helium gas in each of these four lines can be independently adjusted to optimize bearing performance and control spin frequency.  Remotely adjustable mass flow controllers in each gas line are operated as part of a digital feedback loop to maintain any desired rotational frequency below about 5 Hz.  
\subsubsection{The Optical Encoder}
\label{encoder}
Accurately and precisely controlling the rotor spin frequency, of course, requires accurate and precise measurement of that frequency.  This measurement is done using an optical encoder.  The optical encoder is used to measure the angular frequency and position of the drive mass so that it can be phase-correlated with the force signal observed by the interferometer.  The encoder is roughly diagrammed in Fig.\ \ref{spinspeedfeedback}.  It consists of a fiber-optic bundle displacement sensor \cite{spinspeed1}, fit into the lid of the gas bearing, that is aimed at the gold pattern deposited at the edge of the drive mass (see Fig.\ \ref{drivemass} for a photograph of this pattern).  The displacement sensor comprises a bundle of several thousand fiber optic light guides running from the drive mass up the length of the probe and out of the dewar.  The light guides are split into two bunches after they exit the dewar--- one bunch leads to the incandescent lamp that serves as a light source for the sensor, and the other leads to a photodetector.  

The signal from the fiber bundle sensor can be used to measure three quantities: the angular frequency, angular phase, and vertical position of the drive mass rotor.  The angular frequency and phase determine the force applied to the cantilever by any mass-dependent interaction, and the vertical position of the rotor sets the distance between test mass and drive mass. Knowledge of all three of these parameters is thus crucial to the experiment; their measurement proceeds as follows.   

The light from the lamp travels down half of the light guides into the probe, reflects off the drive mass, and re-enters the fiber bundle.  Since the encoder pattern consists of a series of alternating high and low reflectivity areas, rotation of the rotor will modulate the measured reflected power and produce a signal like that depicted in the top trace of Fig.\ \ref{phasedetermination}.  This modulated signal is fed to a data acquisition board, read into a computer, and analyzed using a zero-crossing technique to precisely determine the rotational frequency and phase.  The sense of rotation (clockwise or counter-clockwise) can also be determined, due to the asymmetry of the encoder pattern.

In addition to depending on the reflectivity, the measured reflected power also depends in a non-monotonic but well-characterized way on the absolute distance between the end of the fiber bundle and the reflecting surface.  This position sensitivity allows the fiber bundle sensor to be used to detect the vertical position of the rotor.  In practice, displacements of a few tenths of a micron are easy to detect.  To measure the distance between the drive mass and the cantilever wafer, the experimenter can simply compare the encoder signal when the rotor is spinning to the encoder signal when the rotor is pressed against the lid of the bearing.  The difference between these two signals, converted to a distance, is the separation between the bottom of the wafer and the top of the drive mass when the rotor is spinning.  Since the distance between the cantilever and the bottom of the wafer is fixed by construction and known, this provides an exact and absolute measurement of the all-important face-to-face separation between the drive and test masses.  Calibration of this measurement technique is performed by translating the rigid fiber bundle with a micrometer.
\begin{figure}[tbh!]\begin{center}
\includegraphics[width=0.9 \columnwidth]{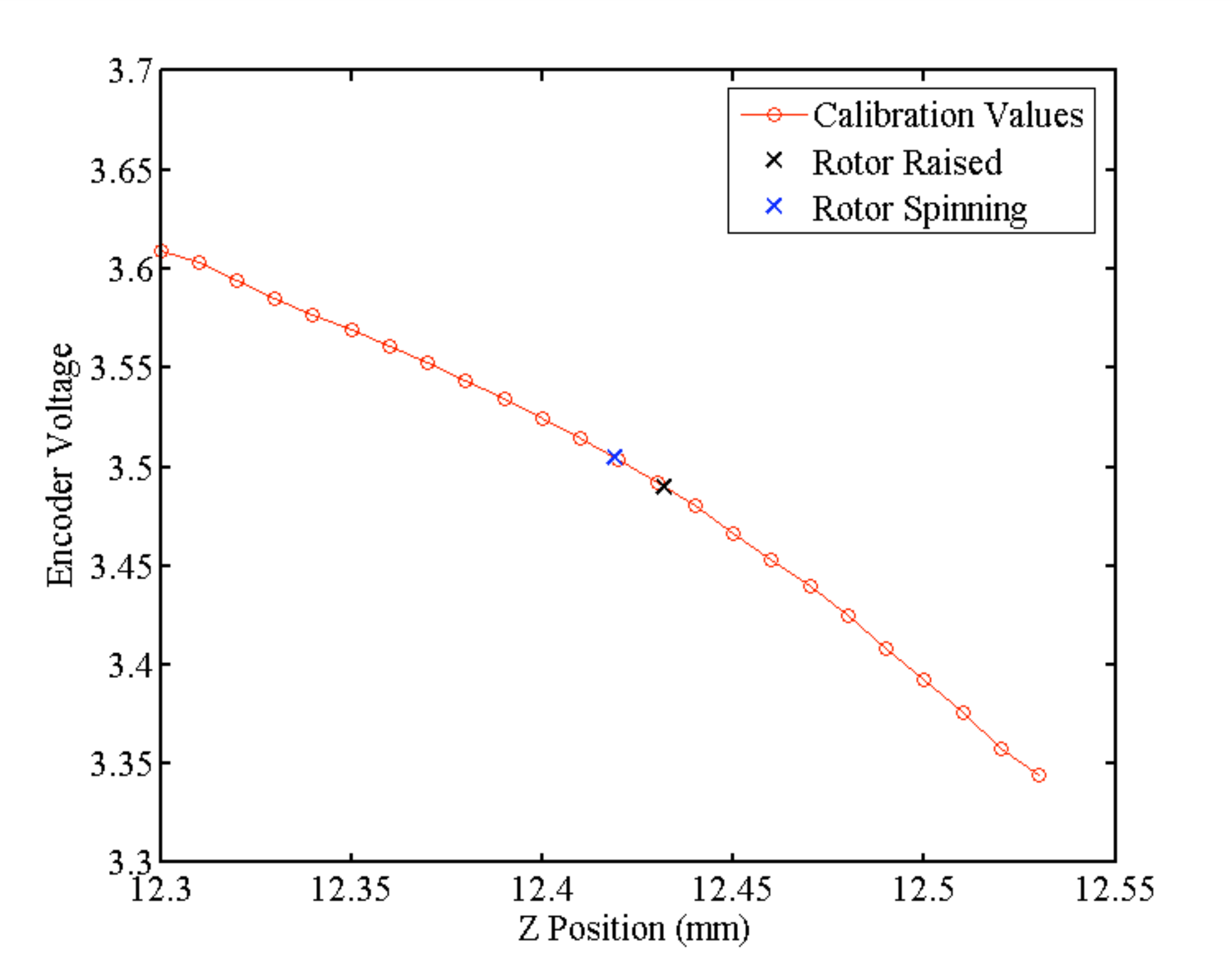}
\caption[Determination of the distance between the masses]{(color online). Data used to determine the equilibrium distance between the top of the housing and the rotor when it is spinning.  The calibration values were taken by moving the fiber bundle with a micrometer.  The ``rotor raised'' reading was taken with the rotor pressed against the lid of the bearing and the fiber bundle aimed at a particular piece of the gold pattern. The ``rotor spinning'' reading was taken while the bundle was pointing at that same gold patch and the rotor was rotating.  The distance between housing and rotor thus determined is 12.8 $\mu$m for this particular measurement.}
\label{zdistance}
\end{center}\end{figure}
The results of such a calibrated $z$-distance measurement are plotted in Fig.\ \ref{zdistance}.  The separation, if the gas flows are properly adjusted, can easily be made as low as 12 $\mu$m.  This wafer-to-drive-mass separation corresponds to about a 29-$\mu$m face-to-face separation between the test mass and the drive mass.
\begin{figure}[tb!]\begin{center}
\includegraphics[width=0.8 \columnwidth]{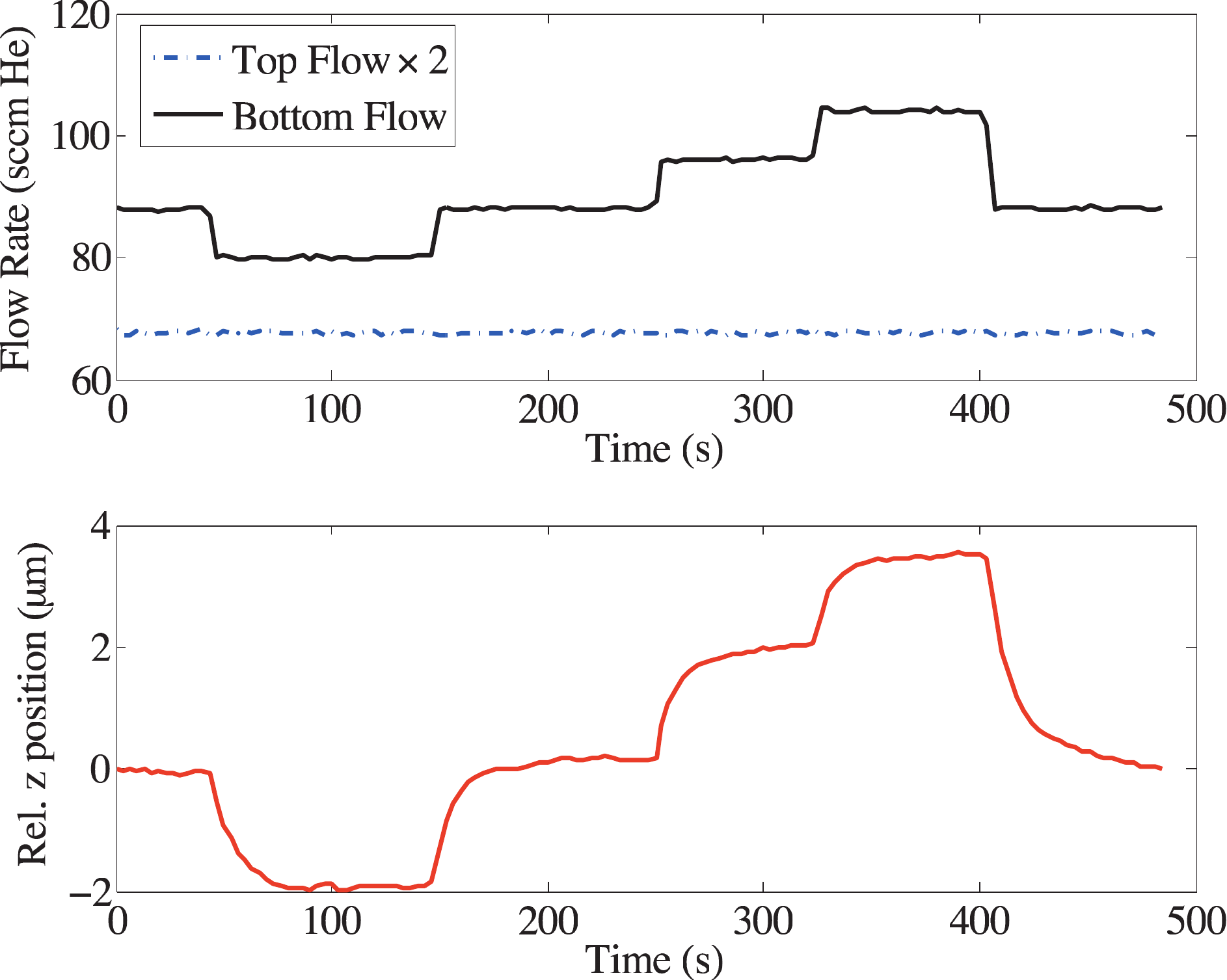}
\caption[Adjusting flows for $z$-distance control.]{(color online). This graph demonstrates the ability to control the $z$-separation over a small range by adjusting the flows in the top and bottom inlets of the gas bearing.  \textbf{Top:} Flow rates versus time, for the top and bottom inlets of the gas bearing.  \textbf{Bottom:} Relative $z$-position versus time, as measured by the calibrated optical encoder.  An overall linear drift, due to thermal contraction of the long rigid fiber, has been removed from the $z$-position data. }
\label{upanddown}
\end{center}\end{figure}
The $z$-distance when the rotor is spinning does depend slightly on the flow rates in the gas bearing (particularly at the top and bottom inlets).  One can thus move the rotor up and down in the cavity while it is spinning by adjusting the flow controllers.  The largest distance by which it is possible to raise or lower the rotor while spinning is about 5 $\mu$m.   An example of the use of this technique is shown in Fig.\ \ref{upanddown}.  
\subsubsection{Feedback Control of the Spin Frequency}
\label{spinfreqFB}
The fiber-bundle-based encoder described above allows very accurate and precise measurement of the rotational frequency of the drive mass.  In order for the experiment to have the maximum possible sensitivity to mass-dependent forces, however, that rotational frequency must also be very tightly \emph{controlled} (not just very well \emph{measured}). 
\begin{figure}[tbh!]\begin{center}
\includegraphics[width=1.0 \columnwidth]{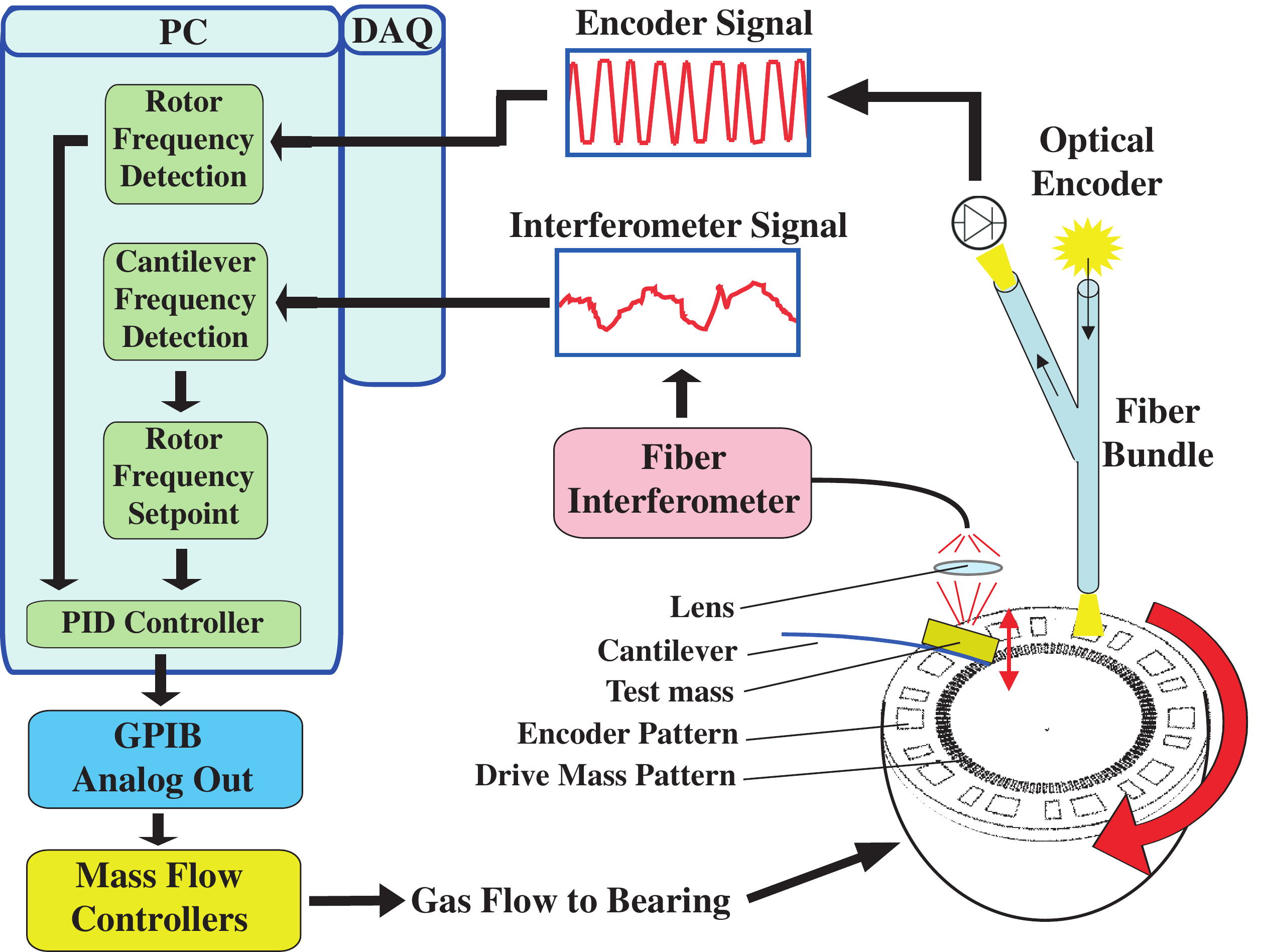}
\caption[Rotor frequency feedback loop]{(color online). A diagram of the feedback loop that controls the spin frequency of the rotor.  The system is described in the text.}
\label{spinspeedfeedback}
\end{center}\end{figure}
A frequency control feedback loop has therefore been implemented, and is schematically represented in Fig.\ \ref{spinspeedfeedback}.  A data acquisition (DAQ) board attached to a PC acquires two signals simultaneously: the signal from the optical encoder, and the signal from the fiber interferometer which is proportional to the cantilever's displacement.  A MATLAB script determines both the rotational frequency of the rotor (by measuring the average time between zero-crossings of the mean-subtracted encoder signal for an integer number of complete rotations) and the resonant frequency of the cantilever (by fitting the Fourier-transformed resonance peak with a Lorentzian).  These values are both sent to a digital PID controller.  The rotor frequency provides the process variable for the PID loop, and the cantilever frequency (divided by 100) provides the setpoint.  The PID controller calculates an error signal based on the difference between the setpoint and the current frequency.  The error signal is added to the current control output, and the new control output is sent to the GPIB-to-analog interface, which passes it on to the mass flow controllers feeding the clockwise and counter-clockwise spin gas inlets.  The control is done in such a way that the total gas flow remains constant while the proportion of the gas going to each spin inlet changes depending on the controller output.  This feedback loop, once tuned, is capable of keeping the spin frequency constant to within 0.5 mHz for several hours.

\subsubsection{Temperature Control of the Gas Bearing}

In addition to controlling the spin frequency of the gas bearing, it is also convenient to be able to control its temperature.  Temperature control is not always required, since the bearing can run stably at base temperature, but the ability to heat the bearing is useful both for the clearing of cryodeposits in the gas lines and for the investigation of thermal dependence of any systematic errors.  In particular, the ability to heat the bearing above the $T_c$ of lead ($\sim$7$^{\circ}$K) provides a powerful discriminator of any effects due to superconductivity in the brass drive mass.

All six gas lines contain heaters which are used for temperature control.  These inline heaters, which sit near the top of the inner vacuum can, consist of wire wound around quartz rods.  The thermometer used for temperature control is a carbon-glass resistor, which is suspended at the mouth of one of the spin gas exhaust ports of the bearing. The thermometer thus measures the temperature of the gas exiting the bearing; at thermal equilibrium this should be the same as the temperature of the spinning rotor, since the rotor is not in contact with anything other than that gas.  Operation of the temperature control is straightforward; the exhaust gas temperature is measured and compared to a setpoint, and the heater currents are adjusted accordingly to increase or decrease the thermal power delivered to the inlet gas.  

\subsection{Drive Masses}

The drive mass consists of an alternating density pattern in a metal disc.  The pattern must be ``buried'' in the following sense: to avoid systematic errors, no variation in the conductivity or height should coincide with the density variation.  We have developed drive mass fabrication procedures for masses made from both brass and tungsten; the brass drive mass was used in our first experimental run. 
\begin{figure}[tbh!]\begin{center}
\includegraphics[width=0.9 \columnwidth]{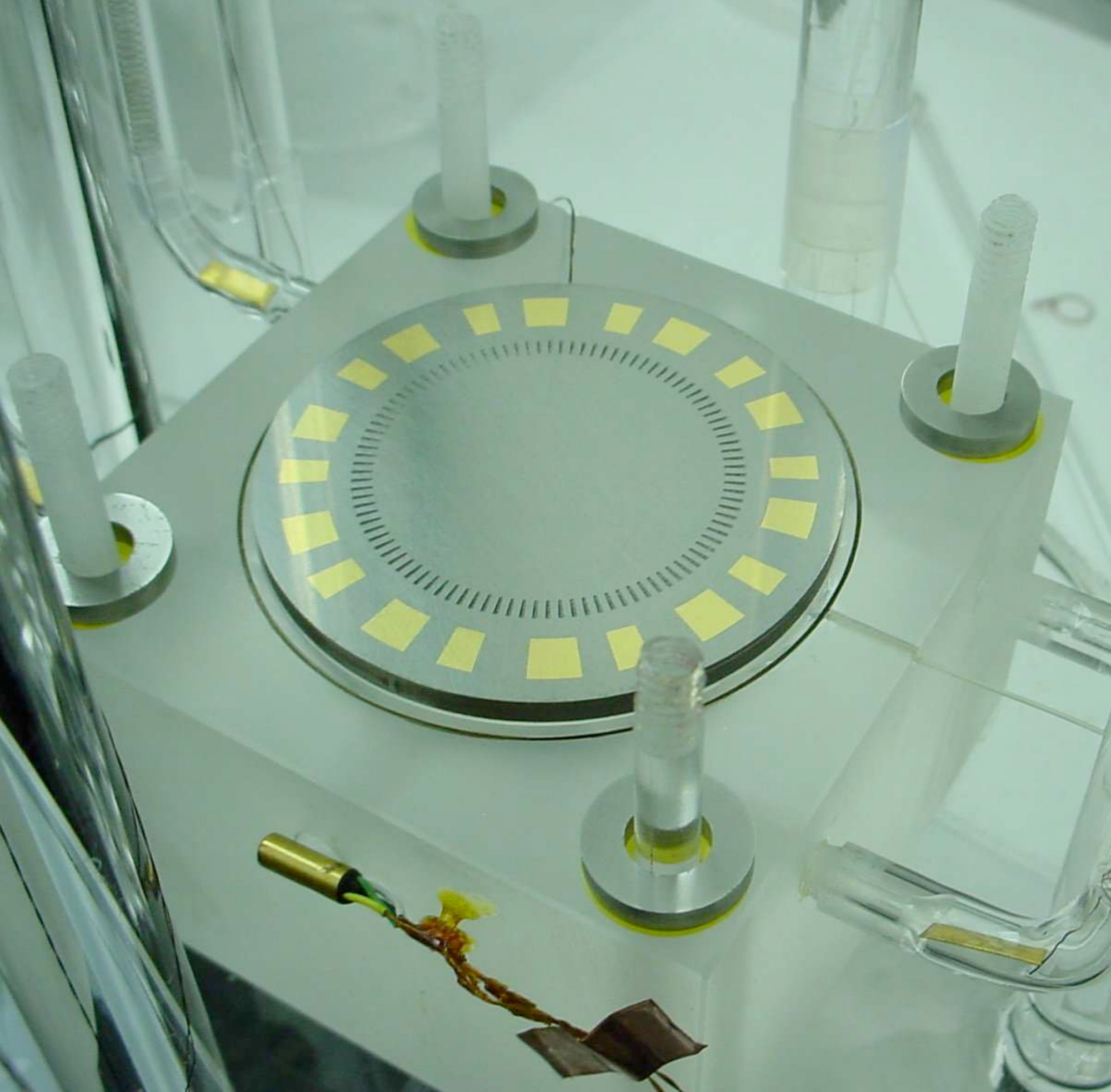}
\caption[Drive mass fabrication]{(color online). A photograph of a drive mass attached to a rotor and installed in the gas bearing.  The drive mass in the photo has not yet been planarized, so the mass trenches are still visible.  Also visible is the gold pattern used by the spin speed detector.}
\label{drivemass}
\end{center}\end{figure}
\subsubsection{Selection of Drive Mass Materials}
Design of drive masses for any experiment that hopes to sense mass-dependent forces must be centered around two goals: maximizing the mass-dependent force, and minimizing the sources of other (mainly electromagnetic) forces.  The first goal can be achieved by using the densest possible material.  Attainment of the second goal, in experiments at this length scale, generally requires the elimination of variations in conductivity, height, and roughness.  A useful feature of this experiment is that drive masses, once constructed, can be switched out relatively easily.  The experiment can thus be run with several different drive mass materials, a capability which is not only a practical advantage but also could prove useful in the investigation of non-Newtonian forces that violate the equivalence principle.

\begin{figure}[tbh!]\begin{center}
\includegraphics[width=0.9 \columnwidth]{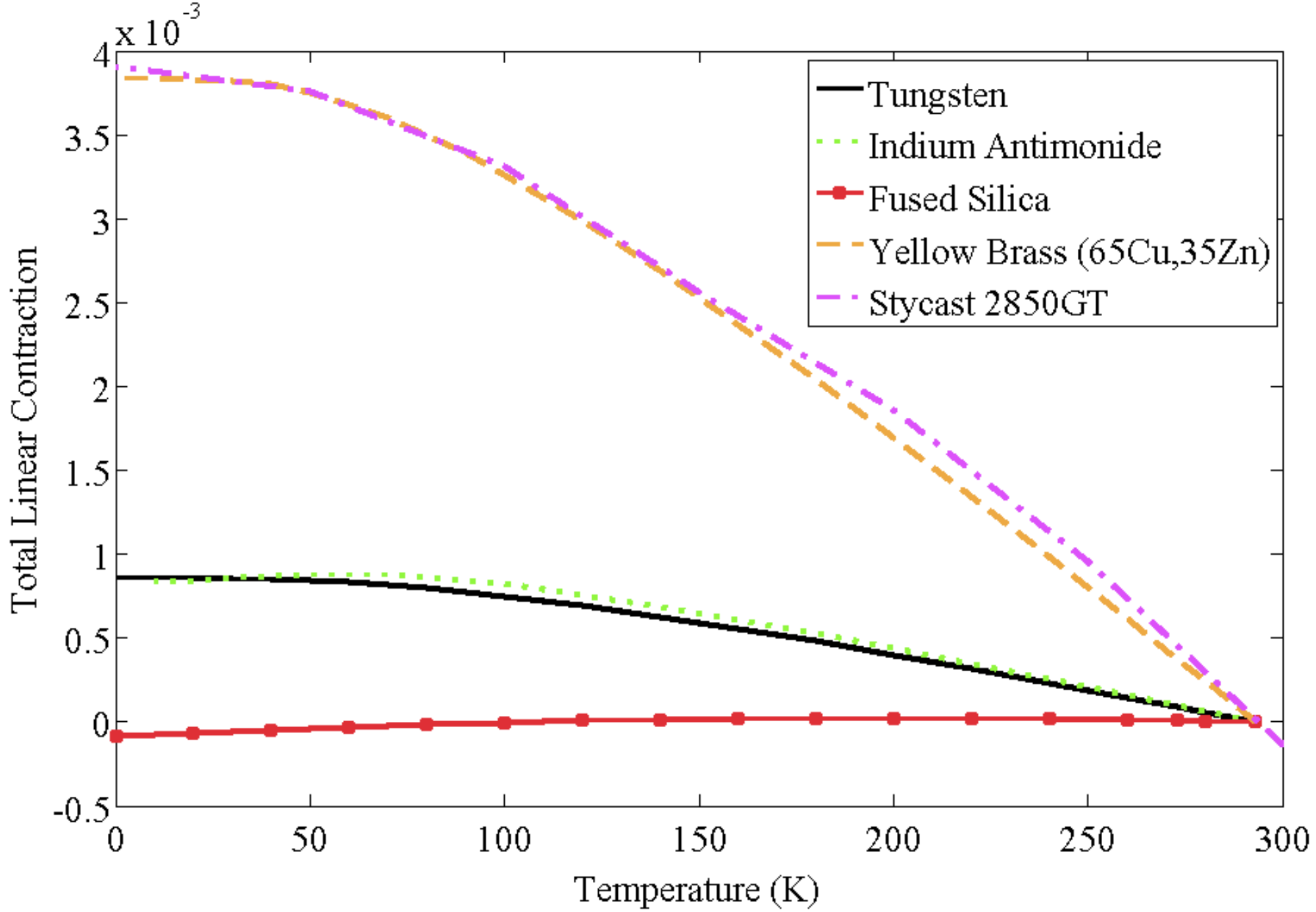}
\caption[Linear contraction of selected drive mass materials]{(color online). The total linear contraction (defined as $\frac{L_{293}-L_{\mathrm{T}}}{L_{293}}$) versus temperature, plotted for selected drive mass candidate materials.  The data are from Refs.\ \cite{thermalexpansionepoxy} and \cite{thermalexpansions}.}
\label{thermalexpansions}
\end{center}\end{figure}

In this experiment our approach has been to use pairs of materials to make the drive mass: a disc made of one dense material, and trenches etched or machined in the disc and filled with another, less-dense material for purposes of planarization.  Since the experiment is designed to be operated at low temperatures, any planarization technique must take account of the fact that most materials contract by a substantial amount when cooled from 300$^{\circ}$K to 4$^{\circ}$K.  If the two materials used to make a drive mass have substantially different linear contractions when cooled, even perfect planarization at room temperature will not result in a mass that is flat at low temperatures. The requirement is thus for pairs of materials with high density contrast but low thermal expansion contrast.  Fig.\ \ref{thermalexpansions} shows some relevant data.
Tungsten immediately suggested itself as a candidate drive mass material--- it is very dense, it is much harder than gold (which eases polishing), it is refractory (which enables some processing steps that require high heat) and it has an anomalously low thermal expansion coefficient for a metal.  The last virtue also poses a challenge, since there are no filled epoxies with thermal expansions low enough to match that of tungsten.  However, somewhat surprisingly, indium antimonide has an almost identical thermal expansion coefficient to that of tungsten, all the way down to 4$^{\circ}$K (see Fig.\ \ref{thermalexpansions}).   Indium antimonide has a low melting point of 525$^{\circ}$C.  This attribute is convenient for drive mass fabrication, as it allows for the possibility of filling the trenches in the tungsten disc with liquid indium antimonide and allowing it to solidify.  

The other drive mass material we have put significant effort into is free-cutting brass.  This material, while substantially less dense than tungsten, has two important advantages: it is vastly more machinable, and its higher coefficient of expansion can be easily matched with commercially available epoxies.  These advantages make it far easier to fabricate a drive mass from brass than from tungsten; indeed, the mass we used in our first experimental run was made with brass and Stycast 2850GT (the epoxy whose linear contraction is graphed in Fig.\ \ref{thermalexpansions}).  Its fabrication is discussed below.  One important feature of the free-cutting brass is a result of the fact that it contains about 2.5\% lead (used to enhance machinability).  This lead is not soluble in the copper-zinc bulk of the brass, and aggregates into macroscopic particles which exhibit superconductivity below about 7.2$^{\circ}$K.   The superconductivity results in a very high magnetic susceptibility; this can lead to a force which is measurable by our experiment, and which can be switched off by heating the drive mass a few degrees.  

\subsubsection{Drive Mass Fabrication}

The fabrication of the drive masses must satisfy the combined requirements of near-perfect planarization, isoelectronic construction, and high density contrast.  We have developed fabrication procedures for two material pairs: tungsten with indium antimonide, and free-cutting brass with Stycast 2850GT epoxy. 

First we discuss the method of fabrication of the tungsten drive mass. A tungsten disc is cut from a plate and a pattern of trenches is machined into it using electrical discharge machining (EDM)---tungsten is too hard and brittle for precise conventional machining.  Since the trenches do not pass entirely through the disc, a ``sinker'' EDM technique using a custom-machined graphite electrode must be applied.  

After the trench machining, the drive mass disc is double-lapped.  The results of the lapping procedure are accurately and precisely checked using an interferometric technique.  Our goal, which we were able to achieve, was to make any flatness errors result in height deviations of less than one micron across the 2-inch drive mass disc.

After polishing and characterizing the flatness of the disc, the trenches must be filled with the thermally-matched companion material indium antimonide.  This is done by heating chunks of indium antimonide placed on the drive mass so that they melt, flow into the trenches, and fill them.  Care must be taken to ensure even heating and good wetting of the indium antimonide to the tungsten.

After the melting step, the drive mass must be lapped to remove the excess indium antimonide that projects above the top surface of the disc.  The result is a tungsten disc with trenches that are filled up to a few microns below the surface of the disc by indium antimonide.  The reason that the top surface of the filler ends up slightly below that of the tungsten is that the softer material is removed more rapidly by the lapping grit. 

When the trenches are satisfactorily filled, the final step of planarization is performed.  This step is the same for both brass and tungsten drive masses, and has been tested more extensively on brass drive masses.  For that reason, we discuss it after the description of brass drive mass fabrication.

The fabrication of the brass drive masses follows essentially the same pattern as that of the tungsten drive masses: initial machining, lapping, trench filling, and planarization.  The particular brass alloy used for the drive masses is free-cutting brass (UNS C36000).  This alloy, which is designed to have excellent machinability, contains approximately 62\% copper, 35\% zinc, and 2.5\% lead.  After machining and lapping the disc, the trenches are filled with Stycast 2850GT, mixed in the appropriate ratio with catalyst 11 (both ingredients are made by Emerson \& Cuming).  This epoxy is designed to match the thermal contraction of brass, as is evident in Fig.\ \ref{thermalexpansions}.  The epoxy is degassed in a vacuum system both after mixing and after potting, to ensure a void-free filling of the trenches.  It is then cured at 90$^{\circ}$C for 24 hours.  As with the indium antimonide, excess epoxy can then be removed by gentle lapping.  After lapping, the top surface of the epoxy is typically one or two microns lower than the top surface of the brass; this is again because of the faster lapping of the filler material.

The filled and lapped drive mass must then be planarized. The planarization procedure is  illustrated in Fig.\ \ref{drivemass}.  The first planarization step consists of covering the whole top surface of the drive mass with Stycast 1266 (also an Emerson \& Cuming product), a very inviscid (650 cP) and slow-curing (8-16 hrs) epoxy.  This epoxy must be made to cure in a very thin and flat layer, so that it will planarize the disc without increasing the distance between the metal of the drive mass and the test mass.  This goal can be achieved by pressing hard on the epoxy while it cures with something that is very flat.  Empirical and theoretical \cite{epoxythickness} analyses agree that the epoxy layer can be made less than a micron thick with a few pounds of pressure, given the stated viscosity and cure time.  The pressure is typically supplied by a large lead brick atop an optical flat.  A layer of smooth aluminum foil (4-mil ``special bright'' from All Foils, Inc.) between the optical flat and the drive mass prevents permanent adhesion of the epoxy to the optical flat, and can be removed without damaging the epoxy layer.

\begin{figure}[tbh!]\begin{center}
\includegraphics[height=1.4in]{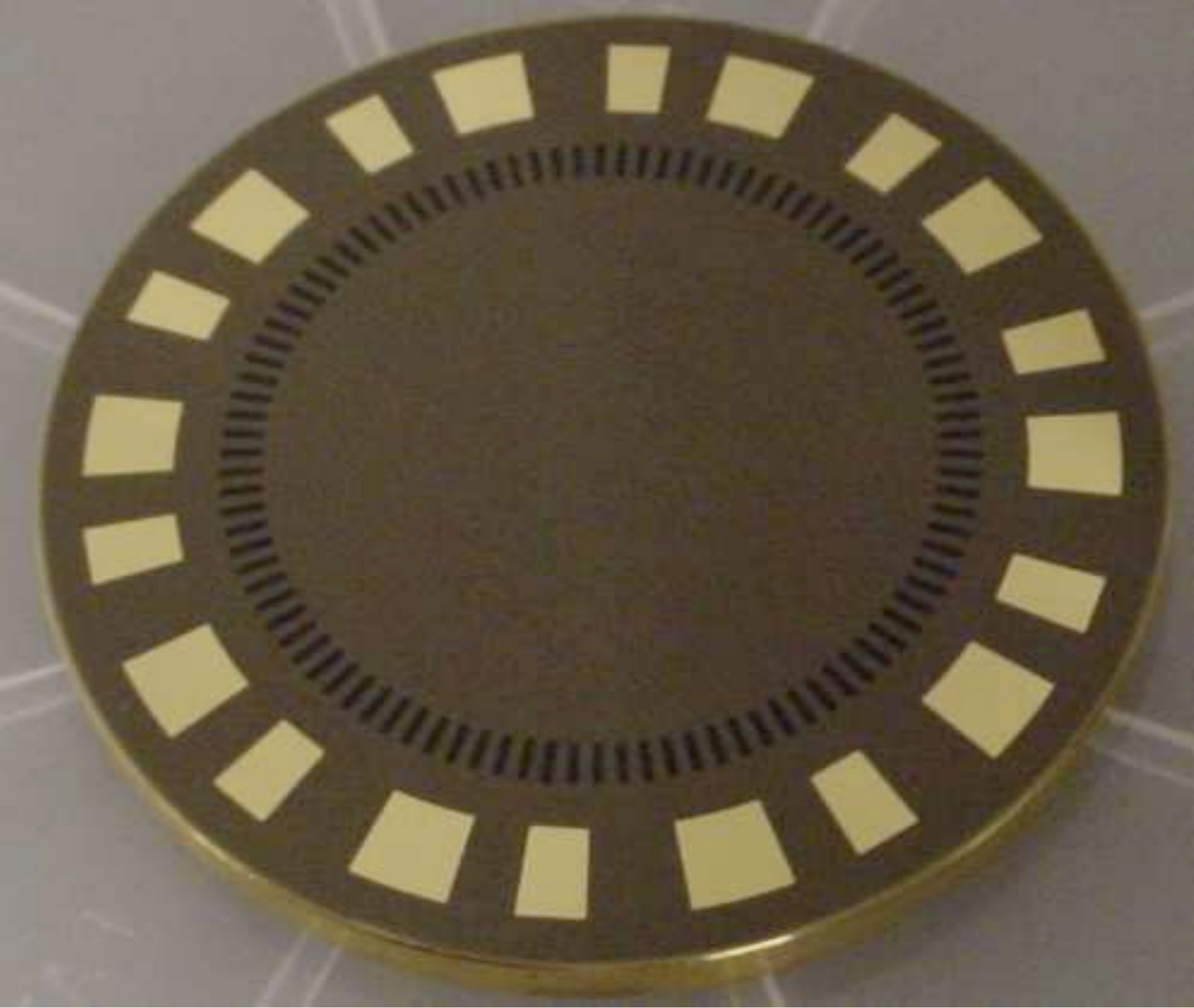}
\includegraphics[height=1.4in]{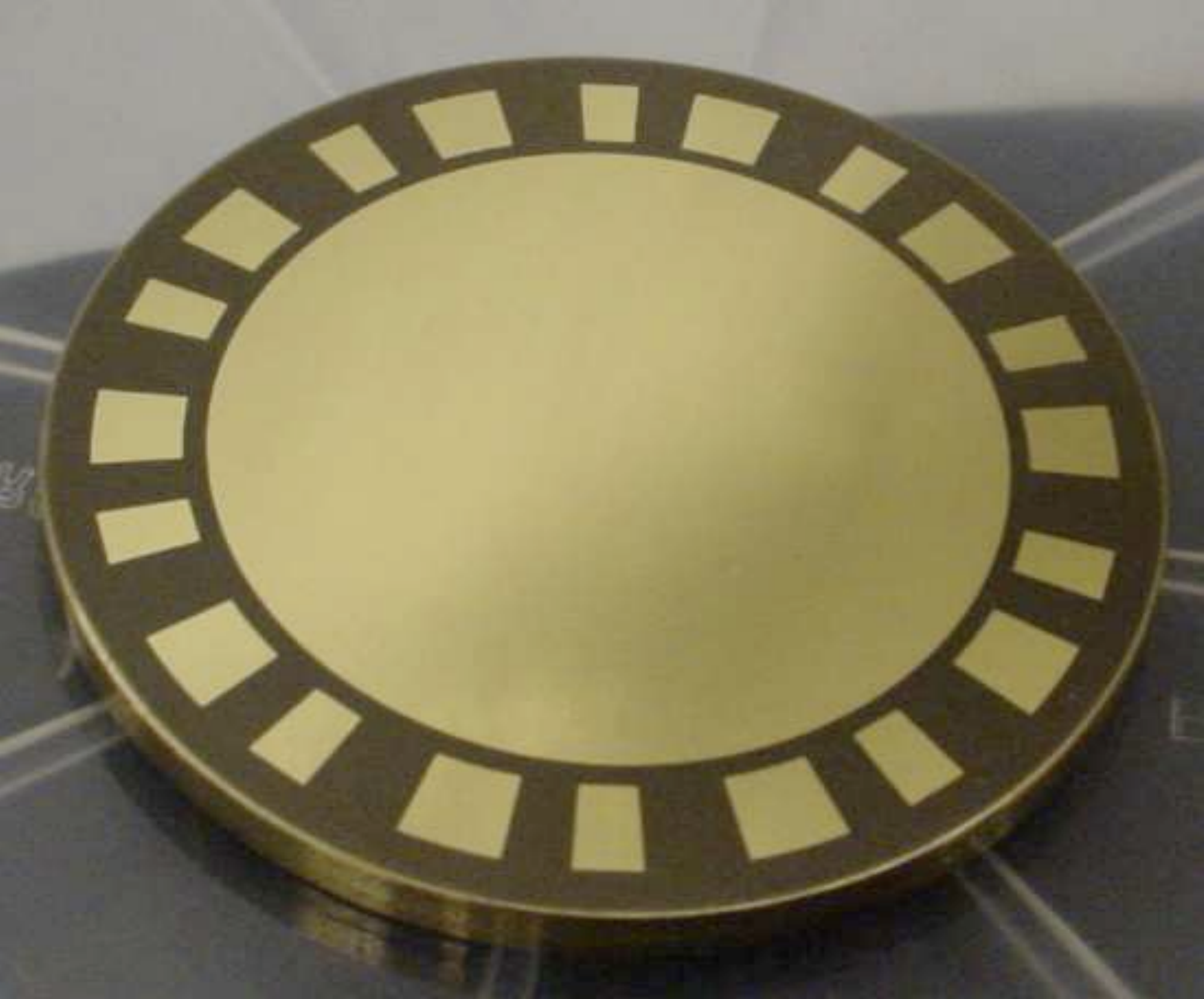}
\caption[Brass drive mass with and without a shield]{(color online). \textbf{Left:} The brass drive mass after final planarization with Stycast 1266.  The black epoxy filling the trenches can still be seen, because the 1266 is optically transparent.  \textbf{Right:} the same mass after deposition of the gold shielding layer.  Note that the mass pattern is no longer visible.  The patterned layer around the edge in both images is part of the optical encoder that detects the disc's rotation.}
\label{brassmassshield}
\end{center}\end{figure}

The result of the final planarization procedure is a flat drive mass disc with a thin ($<1 \mu$m) layer of Stycast 1266 covering the top surface and smoothing out the slight corrugation that remained after the trench-filling.  A 100-Angstrom layer of titanium (for wetting) and a several-thousand-Angstrom layer of gold are then deposited on the disc in a circle that covers the trench pattern.  This layer is what makes the drive mass ``isoelectronic''--- along with the similar layer of gold on the bottom surface of the cantilever wafer, it provides essential shielding against electrostatic and Casimir \cite{casimir} effects.  The brass drive mass is pictured before and after the deposition of this shield layer in Fig.\ \ref{brassmassshield}.

The drive mass is attached to the rotor with a small amount of Stycast 1266 epoxy.  Each drive mass disc must also be accompanied by washers made of the same material and lapped to the same thickness.  These washers support the lid of the gas bearing, and maintain the proper clearance between the drive mass and the lid.  The washers for one of the tungsten drive masses can be seen installed in the probe in Fig.\ \ref{drivemass}.
\subsubsection{Drive Mass Characteristics}
\label{dmcharacteristics}
The end product of this fabrication process is a disc with a circumferential pattern of density variations (100 trenches total) but with no periodic variation in the height or conductivity that might give rise to non-mass-dependent forces.  The density pattern is positioned so that when the drive mass is rotated in the gas bearing, the test mass on the cantilever is alternately above high-density and low-density areas.  Any force that couples to mass will thus produce an oscillating deflection of the cantilever and a time-varying signal in the interferometer, at 100 times the rotational frequency of the drive mass.  This drive mass configuration, with no variation in the conductivity or height, is sometimes referred to as a ``buried drive mass'' \cite{sylviaslac} or ``isoelectronic'' \cite{fischbachexpt} configuration.  Such an arrangement is essential for precision gravity measurements, since the force from gravitational interactions is much smaller than typical electrostatic or even Casimir forces \cite{casimir} at these length scales.  

\begin{figure}[tbh!]\begin{center}
\includegraphics[width=  .95 \columnwidth]{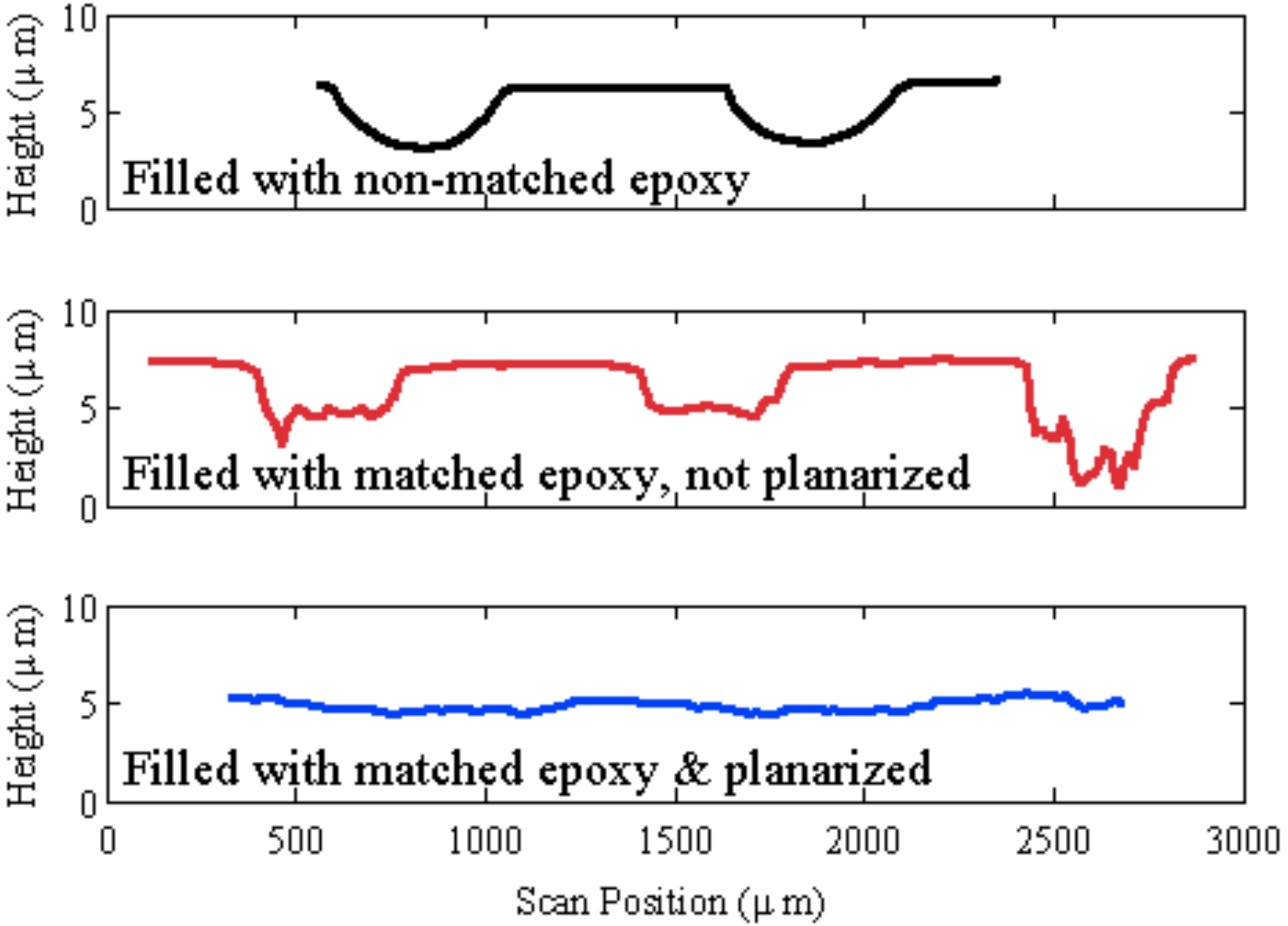}
\caption[Profilometer scans of drive masses]{(color online). Profilometer scans of three different drive masses.  The total trench depth before any filling is 1 mm in all cases.  \textbf{Top:} The trenches of this mass were filled with Stycast 1266.  The large dips visible at the trenches are due to shrinkage in the 1266 during curing.  \textbf{Middle:} The trenches of this drive mass were filled with Stycast 2850GT to a level above the top surface of the mass.  The top surface was then ground and polished to remove excess epoxy.  The dips visible at the trenches are due to faster removal of the epoxy than of the drive mass metal during grinding.  \textbf{Bottom:}  This drive mass (the one with which we ultimately took data) was filled, polished, and planarized, in the manner described in the text.  Variations in the height are mainly due to the roughness of the aluminum foil used for press-curing the planarizing epoxy, and are about 750 nm peak-to-peak.}
\label{alphastepfig}
\end{center}\end{figure}

Once the planarized drive mass was constructed, its flatness was characterized by scanning it with a profilometer.  Ideally, all deviations from perfect flatness should be less than one micron in amplitude, and there should be no evidence of the periodicity of the buried mass pattern.  Profilometer scans of our best drive masses (see Fig.\ \ref{alphastepfig}) showed that they attained this ideal.  The brass mass used for the first data-taking runs was flat to better than 1 micron across its 2-inch diameter, and had roughness of about 750 nm.
It was also important to characterize the low-temperature magnetic properties of the drive mass in order to determine the effects of the superconductivity of the lead, and place limits on the size of any possible magnetic coupling.  To measure the low-temperature magnetic properties, a small piece of one of the brass drive masses was placed in a commercial cryogenic SQUID magnetometer. A superconducting transition of the brass was observed slightly below the bulk critical temperature of lead (7.2$^\circ$K).  Magnetic forces resulting from this superconducting transition can provide a useful check on the force measurement apparatus.  Measurements of gravity with the brass drive mass must be taken with the bearing at an equilibrium temperature above the $T_c$ of bulk lead to avoid spurious signals due to these forces.

\subsection{Test masses}

\begin{figure}[tbh!]\begin{center}
\includegraphics[height=140pt]{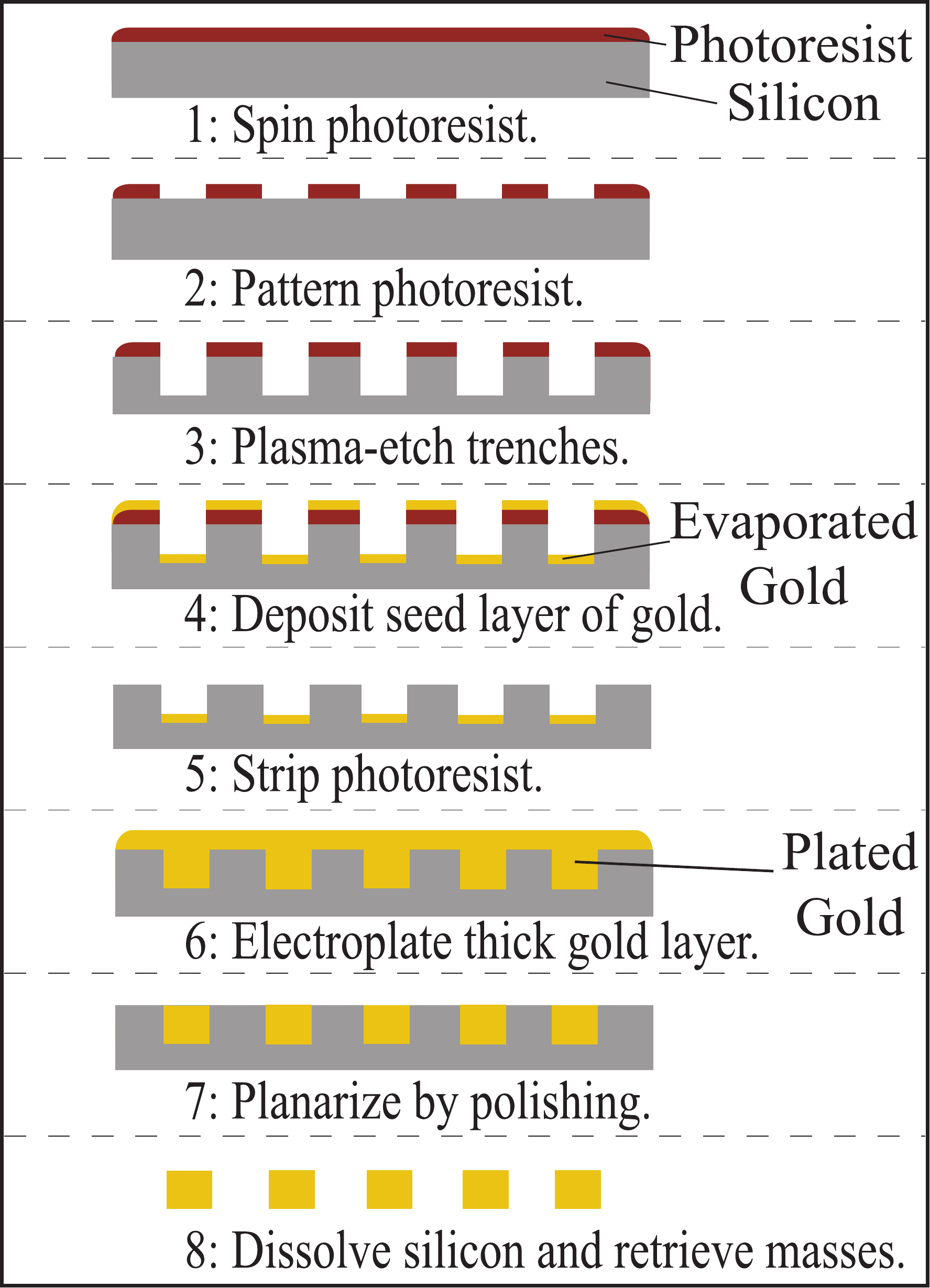}
\includegraphics[height=140pt]{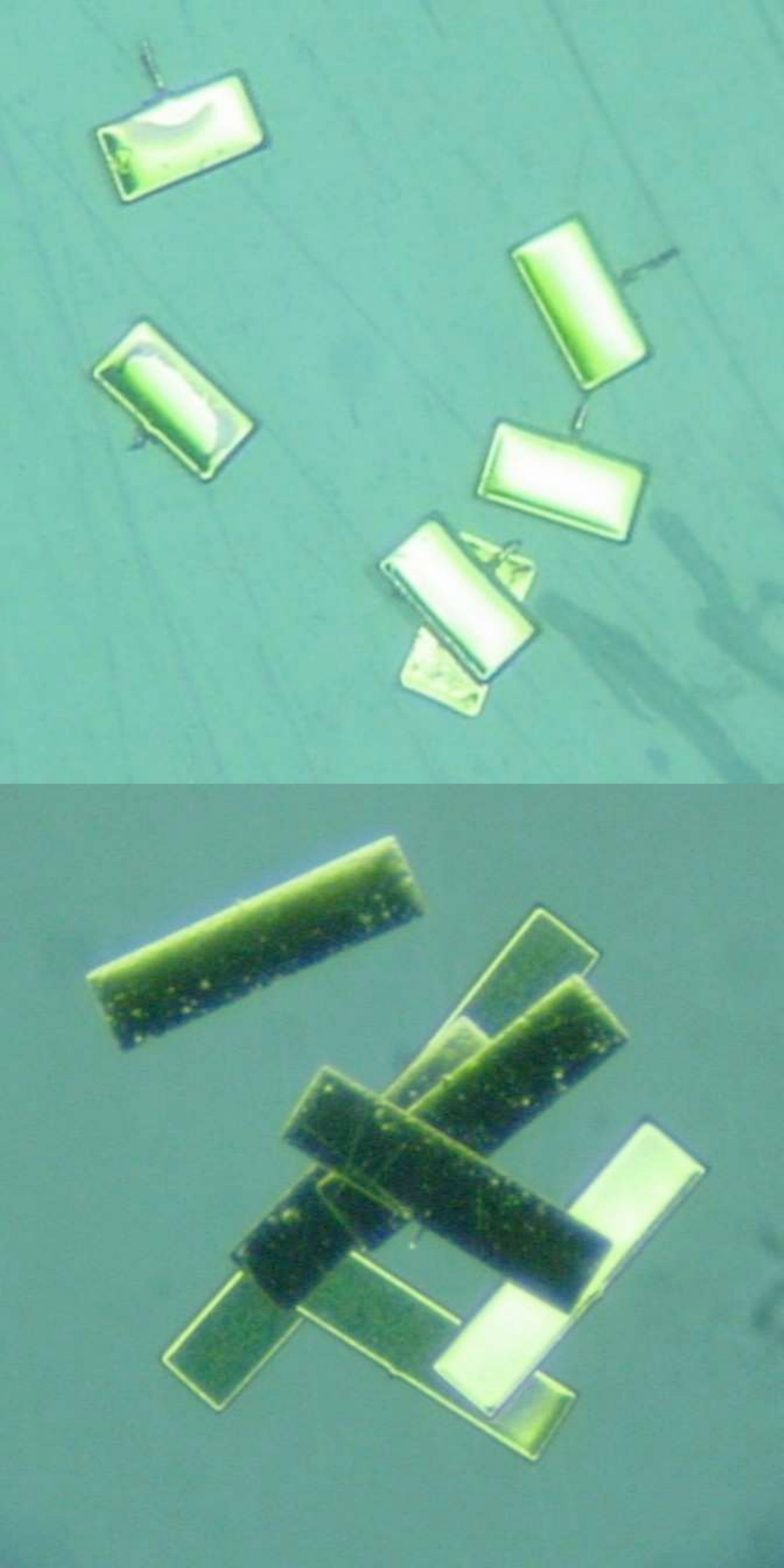}
\includegraphics[height=140pt]{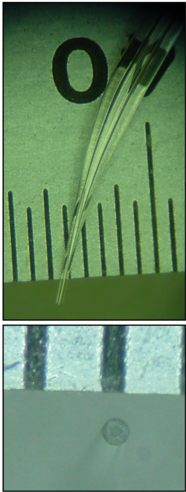}
\caption[Test mass fabrication]{(color online). \textbf{Left:} an outline of the test mass fabrication process, as discussed in the text.  \textbf{Center:} photographs of two different sizes of test mass---200~$\mu$m by 100~$\mu$m (top center) and 400~$\mu$m by 100~$\mu$m (bottom center).  Stubs of the wires used for electroplating are visible on some of the masses. \textbf{Right:} photographs of the glass vacuum chuck used for test mass manipulation and attachment. The $\sim$75~$\mu$m-diameter inlet is visible in the bottom right-hand picture. The lines on the ruler are 500~$\mu$m apart.}
\label{testmass}
\end{center}\end{figure}

The test mass is a small rectangular prism of solid gold a few tens of $\mu$m thick. It is surprisingly difficult to precisely fabricate metallic objects at this length scale, which falls in between the natural length scales of conventional machining and of micromachining.  The best fabrication method for our experiment, an electrochemical micro-casting process, is illustrated in Fig.\ \ref{testmass}.
\subsubsection{Test Mass Fabrication}
First, trenches are plasma-etched into a silicon wafer; the trench pattern consists of an array of small rectangles (200 or 400 $\mu$m by 100 $\mu$m) connected by much smaller trenches 10 $\mu$m wide, and is created by standard lithography using photoresist.    A critical next step not shown in Fig.\ \ref{testmass} is the deposition of a silicon nitride layer.  This step is important because the relatively low resistivity of the silicon wafer can spoil the electroplating process by introducing additional current paths that result in gold deposition in places other than the mass trenches.  The silicon nitride layer has a high resistance and thus eliminates this problem.  A seed layer of gold is deposited at the bottom of the trenches.  Much more gold is then electroplated onto the wafer up to a thickness of 40 or 50 $\mu$m.  The gold in the rectangular trenches will become the test masses, and the smaller trenches are used as wires to make electrical contact during the plating process.  A pulsed-current plating technique is used to avoid problems associated with the buildup of gas at the plating surface and local ionic depletion \cite{pulseplating}.  After the plating, the top surface of the wafer is polished flat and the silicon is dissolved in potassium hydroxide.  What is left is a set of small rectangular solid gold prisms 200 or 400 $\mu$m long, 100 $\mu$m wide, and having a thickness determined by the polishing step.

This process works well, and is substantially higher-yield than previous evaporative techniques this group used for making test masses for a different short-range gravity experiment~\cite{frogland1}.  One feature of both these processes is that there can be fairly substantial thickness variation within a single batch of masses, which can lead to substantial mass variations.  The variation is due to the fact that it is difficult to enforce perfect parallelism across the test mass die during the polishing phase of the process.  In practice, we do see a range of test mass thicknesses--- for example, one batch of 400 $\mu$m--long masses ranges in thickness from 5 $\mu$m to 10 $\mu$m.  This range of thicknesses is convenient, since it enables us to choose several masses to stack on a single cantilever to achieve a desired resonant frequency.  

To measure the thickness variation and check the density of the electroplated gold, an electrobalance was used to weigh a representative sample of test masses.  Individual 400 $\mu$m-long test masses were added to the pan of the balance one by one, and the total mass was recorded after each addition.  The mean slope of the resulting mass versus number graph then furnished an estimate of the mean mass value, and the standard deviation of the mass increase per step gave an estimate of the scatter in mass. The mean mass for this batch of 400 $\mu$m test masses was determined to be 5.4 $\mu$g, and the standard deviation was determined to be $\sim2 \mu$g.  This standard deviation is at least twice the expected error of the electrobalance and thus probably is reflective of a polishing ``wedge'' causing varying mass thickness within the batch. The measured mean mass agrees well with the expected mass of solid gold rectangular prisms with a mean thickness of 7 $\mu$m.

The range of thicknesses of our test masses means that it is difficult to precisely determine the total mass that has been put on any given cantilever.  Once they are attached to a cantilever, though, the masses are on a much more sensitive scale than any electrobalance.  The spring constant of the cantilever can either be predicted from knowledge of its dimensions or measured using radiation pressure \cite{radpress_apl}.  Once the spring constant $k$ of the cantilever is known, the mass $m$ of the attached test masses can be easily determined by measuring the resonant frequency $\omega_{\mathrm{o}}$ of the cantilever:  $m=k/\omega_{\mathrm{o}}^2$, since the mass of the cantilever can be neglected.

\subsubsection{Test Mass Manipulation and Attachment}

The manipulation of the test masses presented a problem similar to the problem of fabricating them.  Too small to hold with tweezers but too heavy to pick up with Van der Waals forces, they fall in between two natural length scales of micromanipulation.  The problem of manipulating the test masses was solved by using a tiny handmade glass vacuum chuck.  A glass micropipette was stretched in a flame and then broken and sanded to produce a bent tube that tapered down to a flat annular end with a small inner diameter.  A small vacuum pump was attached to the larger end of the tube, and the tube was mounted on a micrometer stage.  Since the inner diameter of the small end is smaller than the width of the test masses (100 $\mu$m), the tube can then be used to pick up and move around the tiny masses like an ordinary vacuum chuck.  This technique works very well, and allows reliable and accurate positioning of test masses.  The miniature vacuum chuck can be seen in Fig.\ \ref{testmass}.  Using this tool, the procedure of attaching test masses to the cantilevers with epoxy is fairly straightforward.  The thickness of the epoxy layer between individual masses is much less than a micron--- negligible compared to the thickness of even our thinnest test masses.  The cantilever used in the first data run of the experiment had 4 masses attached to it, which drove the resonant frequency down to about 300 Hz.  A cantilever with multiple 400 $\mu$m masses on it is pictured in Fig.\ \ref{cryopump}.

\label{cantileverattachment}

\subsection{Cantilevers}

\begin{figure}[tbh!]\begin{center}
\includegraphics[width=\columnwidth]{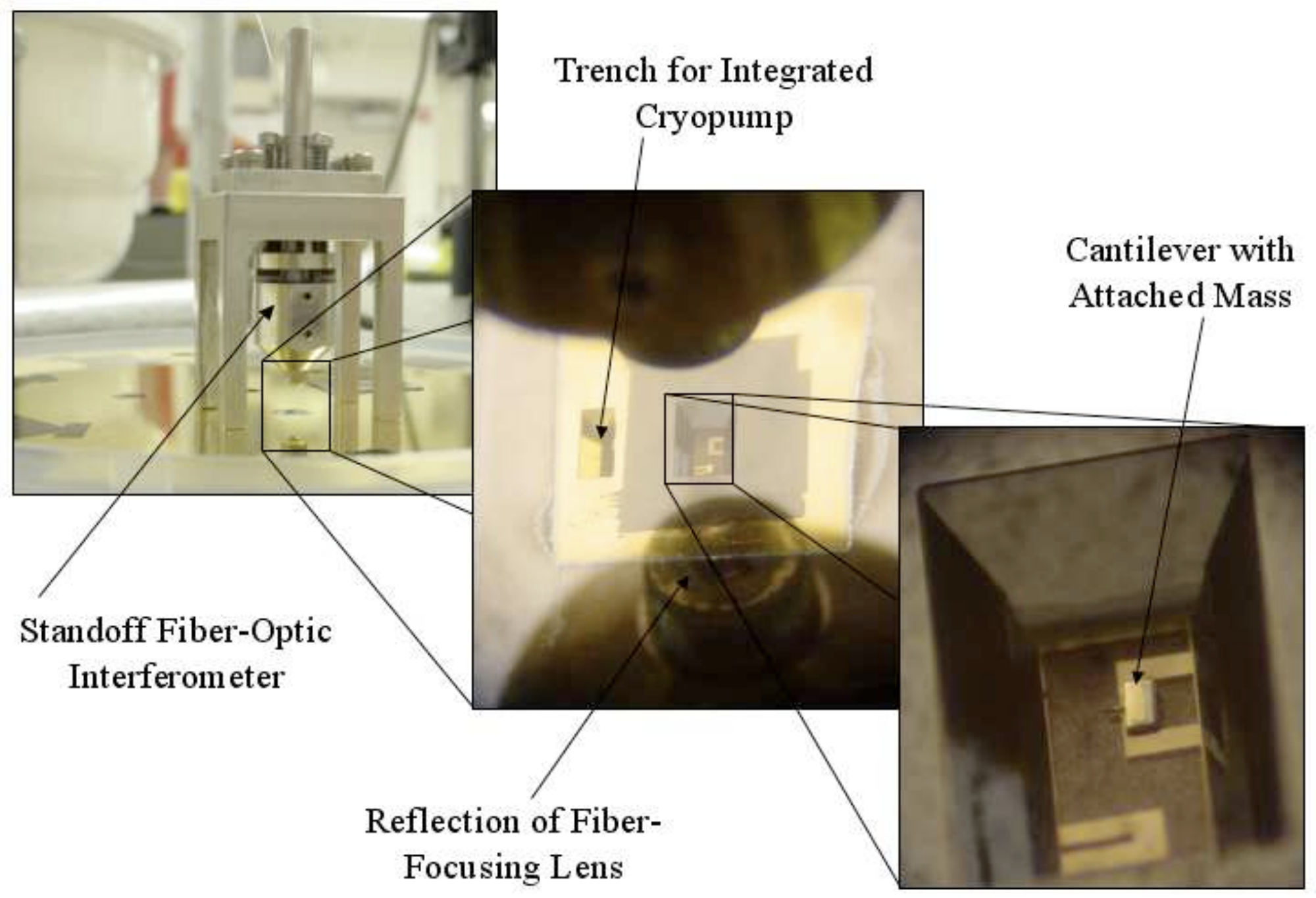}
\caption[Photographs of the cantilevers]{(color online). Photographs of the cantilevers taken at three different magnifications.  Labeled parts are explained in the text.}
\label{cantilevers}
\end{center}\end{figure}

The cantilever is a Hooke's law spring for the small displacements involved in these measurements---it linearly converts the force on the test mass to a displacement.  Our cantilevers have spring constants of about $10^{-2}$ N/m, and resonant frequencies (with the test mass attached) of about 350 Hz.  Photographs of the cantilevers appear in Fig.\ \ref{cantilevers}. 
\subsubsection{Cantilever Design}
The design of the cantilevers was the subject of an extensive optimization process, due to the need to balance many competing factors to maximize the experimental sensitivity.  Some of these factors included the spring constant (lower is more sensitive), the frequency (higher is better for averaging purposes), the width, the length, the quality factor (higher is better), the fabrication yield, and the amount that the mass-loaded cantilevers droop in the earth's gravity (lower is better, for parallelism and to avoid stiction \cite{stiction}).  Ultimately we decided to use pairs of cantilevers in two different sizes.  Both are two-legged cantilevers, rather than rectangular ones--- this design allows a cantilever to take wider test masses while keeping the spring constant low.  

\subsubsection{Cantilever Wafer Fabrication}

\begin{figure}[tbh!]\begin{center}
\includegraphics[width=0.9 \columnwidth]{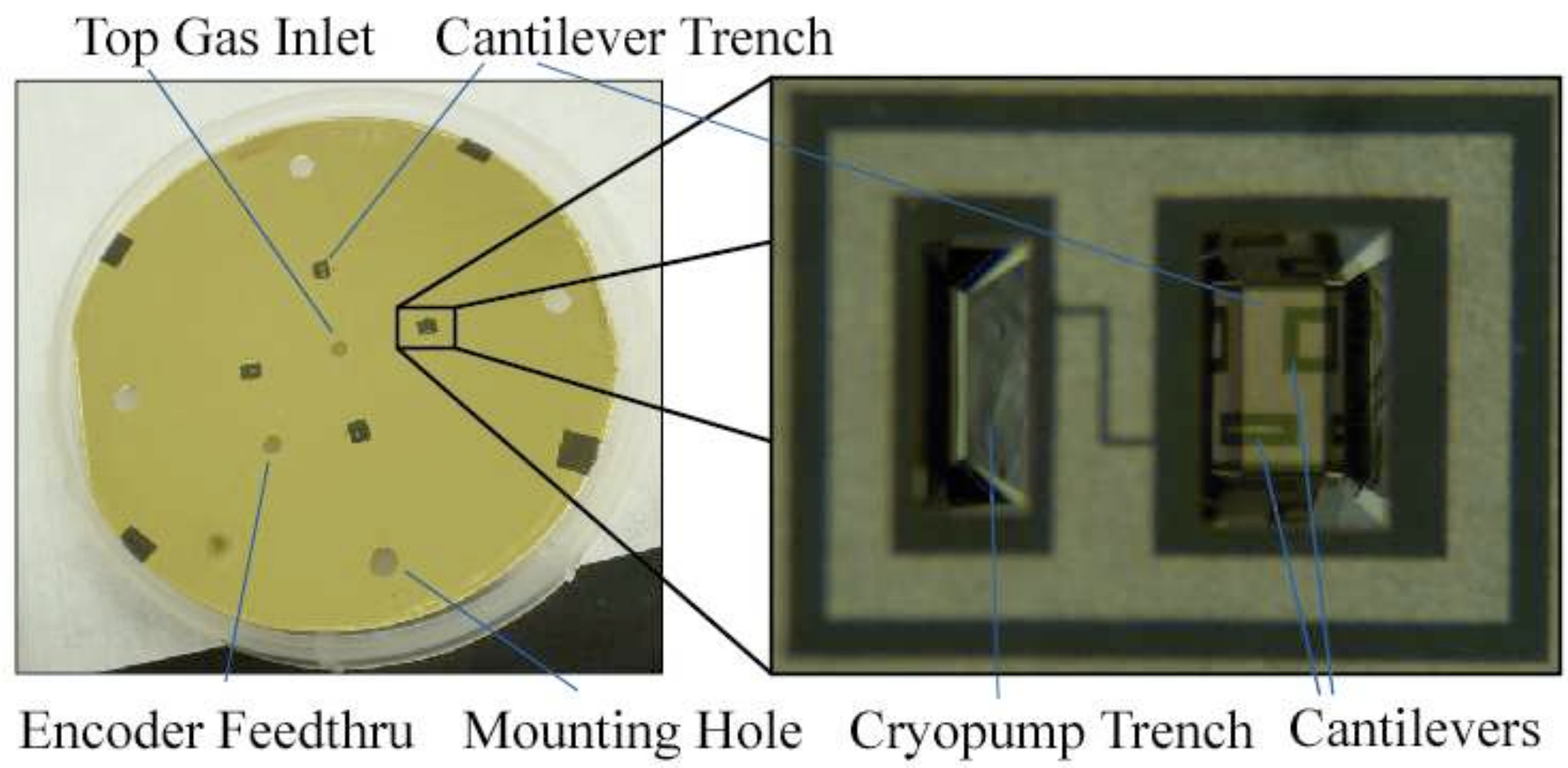}
\caption[Photographs of cantilever wafer]{(color online). \textbf{Left:} a photograph of a whole cantilever wafer.  All of it except the area inside the trenches has been coated with titanium-platinum-gold.  \textbf{Right:} A close-up photograph of the cantilever trench and the cryopump trench.  Reflections from the angled side wall of the cantilever trench, which are visible in the close-up, are of great help during precise positioning of test masses.   Labeled parts are explained in the text.}
\label{cantileverwafer}
\end{center}\end{figure}

We use an un-diced 4-inch cantilever wafer to form the top flat surface of the gas bearing.  The cantilever wafers were fabricated using standard microfabrication techniques.  Each wafer contains eight cantilevers, arranged in pairs about 10 $\mu$m above the bottom of four trenches around the wafer.  Each trench is etched all the way through the silicon wafer to a 4-$\mu$m-thick silicon nitride layer at the bottom of the wafer.  This nitride layer, when coated with gold, forms the shield membrane.  This membrane provides a second layer of electrical shielding between the drive and test masses and allows the cantilevers to be kept in vacuum while gas is flowing in the bearing.  \label{shielddescription}  Six round through-holes are etched in the wafer during fabrication: on in the center for the top bearing gas inlet, one between two cantilever trenches for the optical encoder to look through, and four around the edges for the quartz bolts attaching the bearing lid to pass through.  A cantilever wafer is pictured in Fig.\ \ref{cantileverwafer}.
\subsubsection{Cantilever Displacement Detection}

The cantilever's deflection is measured by a laser interferometer, based on the design presented in Ref.\ \cite{rugar-intf}.  The interferometer uses the optical cavity formed between the reflective surface of the gold test mass and the end of the optical fiber used to inject the light.  A lens inside this cavity focuses the light on the test mass; this allows the cavity to be several millimeters long, so that the optical elements can be positioned outside of the sealed cantilever trench.  The sensitivity of an optical interferometer is a periodic function of $d/\lambda$, where $d$ is the cavity length and $\lambda$ is the wavelength of light being used (1310 nm).  This long optical cavity thus also allows the interferometer to be kept at the point of maximal sensitivity by temperature controlling the laser instead of mechanically adjusting the cavity length.  The necessity for a piezoelectric biasing element in the Fabry-Perot cavity is thus avoided; this both simplifies the design and greatly reduces susceptibility to electrical noise.  This technique requires that the coherence length of the laser be long compared to the cavity, so a high-coherence-length fiber-coupled distributed feedback (DFB) laser is used as a light source for the interferometer.  

Assuming the coherence length is longer than the cavity length, the optical power reflected from the Fabry-Perot cavity depends on the cantilever's position as $P_{\mathrm{out}}=P_{\mathrm{o}} (1-V\cos 4\pi d / \lambda)$, where $\lambda$ is the wavelength of the laser, $d$ is the distance from the cleaved fiber end to the cantilever, $P_{\mathrm{o}}$ is the midpoint power, and $V$ is the interference fringe visibility \cite{rugar-intf}.  The laser light is injected through one arm of a fiber coupler, as shown in Fig.\ \ref{setup}; a photodiode attached to another arm of the coupler produces a current proportional to the reflected power; this current is then converted to a voltage by a transimpedance amplifier.  This voltage is then acquired by the DAQ, and serves as a fast and accurate measurement of the position of the cantilever.

Physically, the fiber interferometer is a subset of the radiation pressure damping apparatus shown in Fig.\ \ref{setup}.  The non-free-space optical elements of the fiber interferometer are the coupler, the photodiodes, and the wave division multiplexer used to split off the 1550 nm light used for damping the cantilever motion.  The only free-space portion of the interferometer is the Fabry-Perot cavity itself. A stainless steel ferrule holder with four legs is glued to the cantilever wafer or to the glass lid of the cantilever trench.  This piece maintains the perpendicularity and alignment of the fiber to the test mass as the probe is cooled to low temperatures.  Clamped into the ferrule holder is a stainless steel ferrule which holds both the glass fiber ferrule and the aspheric lens used for focusing, and maintains their relative coaxial alignment.  
\subsubsection{Integrated Cantilever Vacuum System}
\label{sealingdescription}
An important design challenge in this experiment was that the gas bearing (used as a mass actuation system) and the cantilevers (used as force sensors) require radically different pressure ranges.  The gas bearing exhausts helium gas into the probe space at a pressure as high as a few hundred torr, but the cantilevers in the gas bearing lid need to be in a vacuum substantially better than $10^{-3}$ torr in order to make use of their high intrinsic quality factor.  This latter condition is due to the fact that the gas viscosity dominates cantilever damping above about $10^{-3}$ torr \cite{kevinthesis}.  This design problem was solved by enclosing the cantilevers in a hermetically sealed cavity that was maintained at a much lower pressure than the surrounding environment of the gas bearing.

The gold-coated silicon nitride shield membrane described in section \ref{shielddescription} forms the bottom of the sealed cavity.  The sides of the cavity are the walls of the anisotropically etched trench in which the cantilevers sit.  The lid of the sealed cavity must be added by hand after microfabrication; it needs to be optically transparent, in order to allow interferometric detection of the cantilever's motion.  In order to provide a reasonable thermal expansion match with the silicon wafer, small discs of borosilicate glass were used as lids.  To seal the lid of the cavity, a small gasket of indium wire was placed around the cantilever trench, the glass lid was pressed against the gasket with a small weight, and the assembly was heated above the melting point of indium ($157^{\circ}$C) while under vacuum.  Both the lid and the cantilever wafer were coated with a wetting layer of  80~\AA~of titanium, 100~\AA~of platinum, and 1000~\AA~of gold.  The platinum is necessary to avoid scavenging and de-wetting of the gold by the indium.  This method was suggested by a technique proposed for use in the assembly of laser diodes \cite{indiumsealing}.
\begin{figure}[tbh!]\begin{center}
\includegraphics[width=\columnwidth]{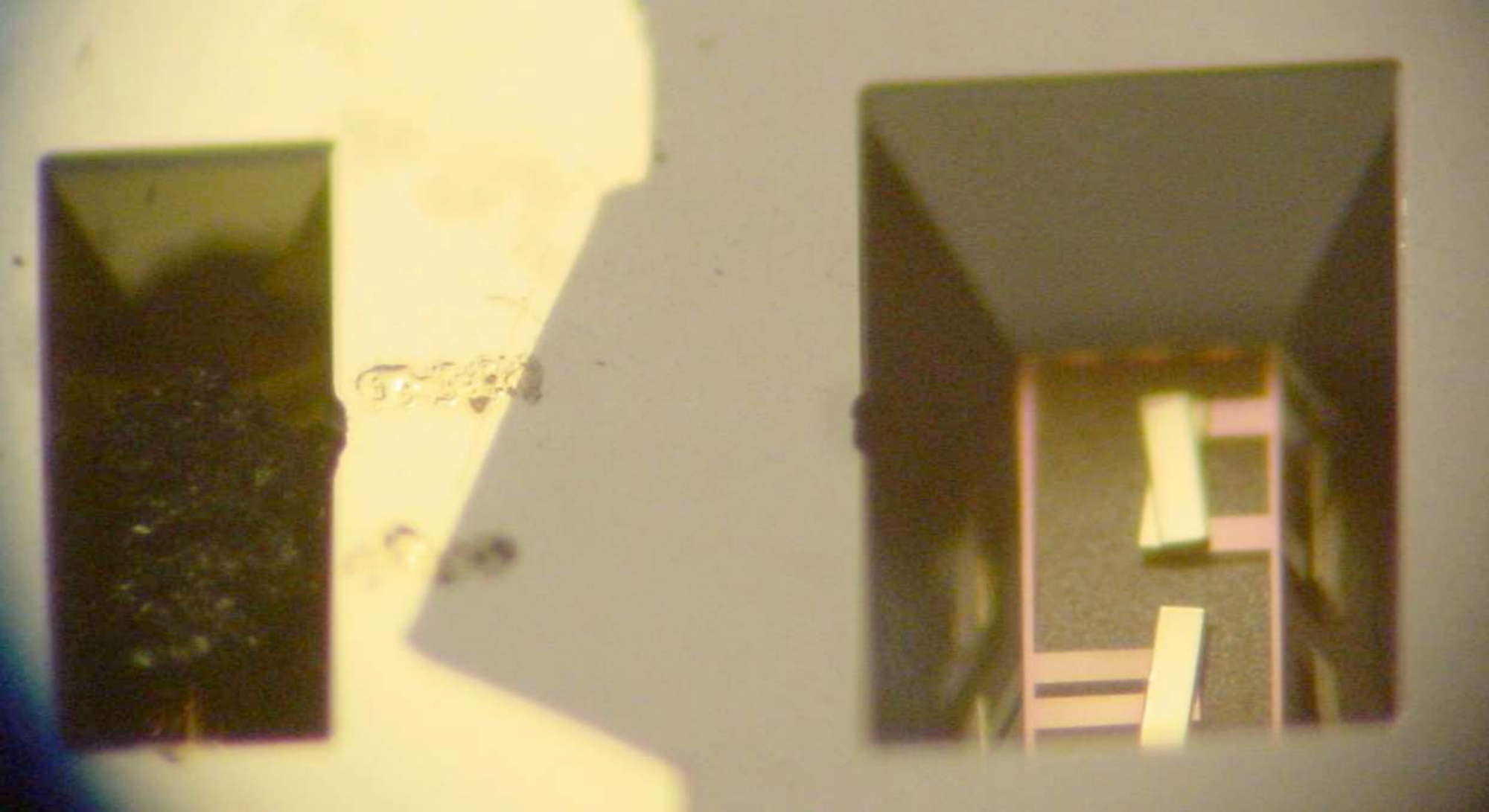}
\caption[Microfabricated cryopump]{(color online). A photograph of the microfabricated cryopump.  The charcoal piece is visible in the trench to the left.  The two smears between the trenches are epoxy bumps, placed there to ensure that gas from the cantilever trench can get to the cryopump trench.  Both cantilevers in the trench to the right have 400 $\mu$m masses on them.}
\label{cryopump}
\end{center}\end{figure}
Even though the sealing of the cavity is performed under vacuum, the measured $Q$ of the sealed-in cantilevers is still fairly low at room temperature (less than 100, as compared with an intrinsic $Q$ of around 80,000).  This low quality factor is a result of viscous damping by residual gases released during the soldering process that are trapped inside the cavity after the seal is made.  This problem is fairly generic, and occurs to some extent with almost all sealing techniques that are commonly used for encapsulation of MEMS devices.  To get around it, one needs to pump on the cavity after it is sealed.  In order to do this, we built a microfabricated cryopump that ensures that the residual gas pressure in the cavity will be negligible at low temperatures.  The miniature cryopump is visible in Fig.\ \ref{cryopump}.  A piece of activated charcoal is epoxied into an auxiliary trench connected to each main cantilever trench.  The charcoal is sealed in with the cantilevers, and the cryopump trench communicates with the cantilever trench either through a microfabricated pumping line (as in Fig.\ \ref{cantileverwafer}) or, more simply, via a gap between the lid and the wafer caused by small dabs of epoxy between the trenches (as in Fig.\ \ref{cryopump}).
\begin{figure}[tbh!]\begin{center}
\includegraphics[width=\columnwidth]{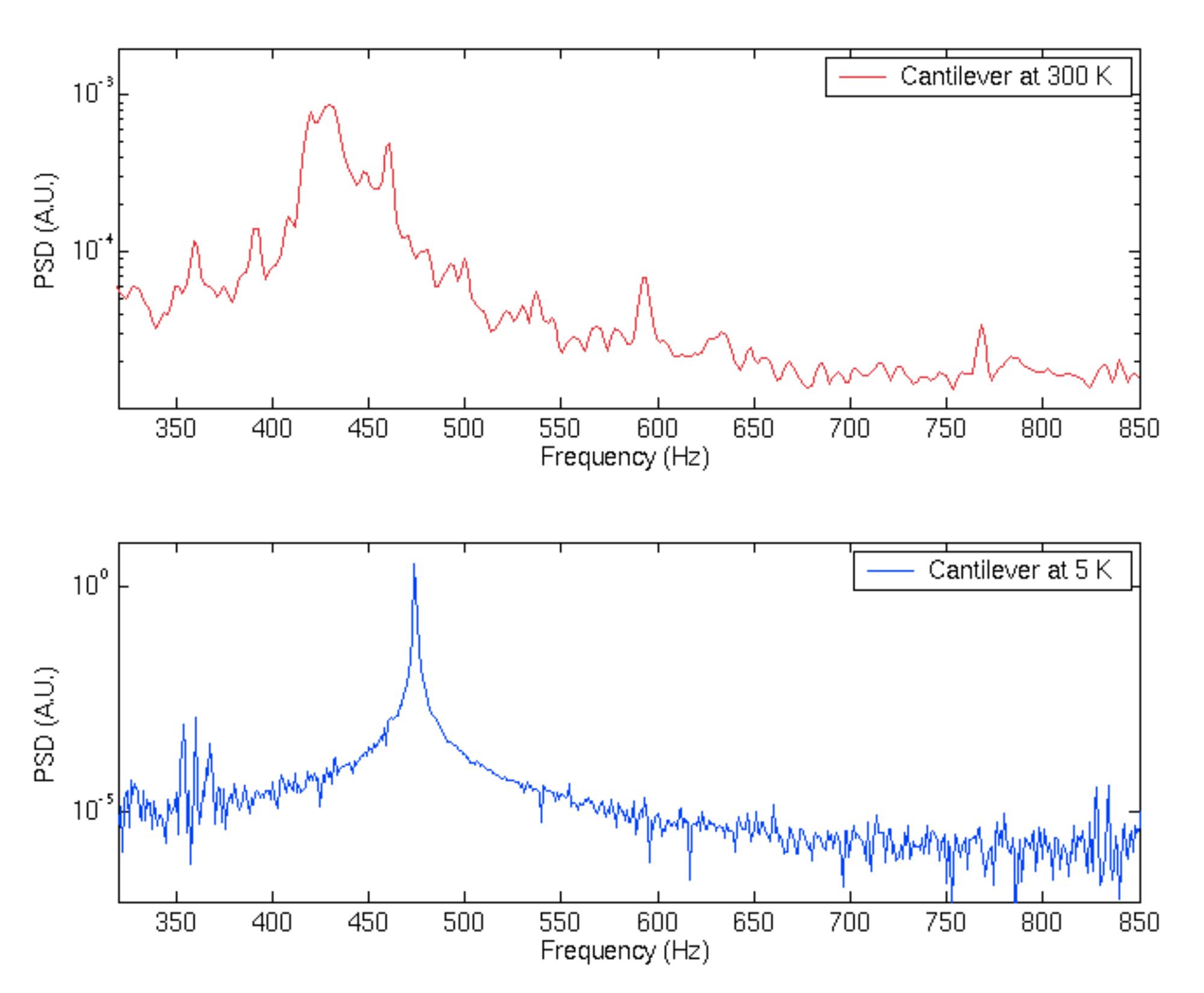}
\caption[Cryopump effectiveness]{(color online). \textbf{Top:} thermal noise spectrum of a cantilever's motion at room temperature.  The cantilever is sealed in a cavity with a microfabricated cryopump, as described in the text.  The width of this resonance corresponds to a quality factor of $\sim$10.  \textbf{Bottom:} thermal noise spectrum of the same cantilever at approximately 5$^{\circ}$K.  The width of this much-sharper resonance corresponds to a quality factor of $\sim$80,000.  This very high $Q$ illustrates the effectiveness of the cryopump, but makes it difficult to operate the rotor on resonance.}
\label{cantileverQ}
\end{center}\end{figure}
When the wafer is cooled to cryogenic temperatures, the large surface area of the charcoal cryopumps out virtually all residual gases left over from the sealing process and produces an excellent vacuum inside the sealed cavity.  This encapsulated vacuum is important for maximizing the quality factor of the cantilevers and thus maximizing the overall force sensitivity of the experiment.  The effectiveness of the microfabricated cryopump is illustrated in Fig.\ \ref{cantileverQ}, which shows that the $Q$ of the cantilever increases by nearly four orders of magnitude when the wafer is cooled to liquid helium temperatures.  This increase is due largely to the removal by the cryopump of virtually all the gas that causes viscous damping of the cantilever's motion at room temperature.

\subsubsection{Radiation Pressure Damping of the Cantilevers}

\label{frequencystabilityproblem}

The sealed and cryopumped cantilever cavity described in the previous section results in a very high cantilever quality factor at low temperatures.  This very high $Q$ is good in the sense that it improves force sensitivity, but it causes an important practical difficulty, which we address here.  The problem is that the width of the resonance of a cantilever with a quality factor $Q$ of 80,000 and a resonant frequency $f_{\mathrm{o}}$ of 350 Hz is equal to $f_{\mathrm{o}}/Q$, or about 5 mHz.  In order for the excitation of the cantilever due to the drive mass to take place on resonance, the excitation frequency must not wander from the resonant frequency by more than this width.  Since there are 100 drive mass patterns per circumference of the drive mass disc, this means that the rotation frequency of the gas bearing rotor must be controlled to within $(f_{\mathrm{o}}/Q)/100$, or about 50 $\mu$Hz.  Unfortunately, it is not practically possible to control the rotation of the gas bearing with such precision.    

The solution to this problem is to reduce the $Q$ of the cantilever, ideally without compromising the good force sensitivity that is a result of the high intrinsic $Q$.  This can be done by feeding back a phase-shifted version of the cantilever's displacement to the cantilever as a force.  To that end, we developed a simple method for feedback-regulation of the response of a microcantilever using the radiation pressure of a laser.  That method is described briefly here, and in detail in Ref.\ \cite{radpress_apl}.

The displacement $x$ of a damped harmonic oscillator as a function of frequency $\omega$  is 
\begin{equation}
x(\omega) = \frac{\omega_{\mathrm{o}}^2/k}{\omega_{\mathrm{o}}^2 - \omega^2 + i\Gamma\omega} [F_{\mathrm{thermal}}(\omega) + F_{\mathrm{ext}}(\omega)],
\end{equation}
where $k$ is the spring constant, $\omega_{\mathrm{o}}$ is the resonant frequency, and $\Gamma=\omega_{\mathrm{o}}/Q_{\mathrm{o}}$ is the intrinsic damping of the oscillator.  Here, $F_{\mathrm{thermal}}(\omega)$ represents the random thermal Langevin force and $F_{\mathrm{ext}}(\omega)$ an externally applied force, which in this case is due to radiation pressure.  The applied force can be modulated by a feedback loop whose input is the measured displacement.  Adjusting the phase of the feedback gain at the resonant frequency to $\pi/2$ has the effect of producing a velocity-dependent force at the resonant frequency.  In particular, if the gain is chosen so that the applied force near resonance is
$F_{\mathrm{ext}} = -i m \omega g  x$,
where $m$ is the mass of the oscillator and $g$ is proportional to the magnitude of the feedback gain on resonance, then the displacement as a function of frequency becomes
\begin{equation}
x'(\omega) = \frac{\omega_{\mathrm{o}}^2/k}{\omega_{\mathrm{o}}^2 - \omega^2 + i[\Gamma+ g]\omega} [F_{\mathrm{thermal}}(\omega)].
\end{equation}
Assuming that the noise of the feedback system can be neglected, the feedback thus changes the damping of the system without adding fluctuations.  This changed damping leads to a changed effective quality factor $Q_{\mathrm{eff}}=\omega_{\mathrm{o}}/[\Gamma + g]$ and a changed effective temperature $T_{\mathrm{eff}}=T_{\mathrm{o}} \Gamma/[\Gamma + g]$, where $T_{\mathrm{o}}$ is the temperature of the oscillator's environment \cite{cohadon}.  A positive $g$ lowers both $Q$ and $T$ by the same factor.

In order to realize a setup for using radiation pressure to control the $Q$ of the cantilever, our fiber-optic interferometer was modified so that it uses one laser to read out the position of the cantilever and another laser of a different wavelength to apply a force that is a phase-shifted function of that position.  At room temperature, we were able to achieve a reduction of both $Q_{\mathrm{eff}}$ and $T_{\mathrm{eff}}$ of our silicon nitride cantilever by more than a factor of 15 using a root-mean-square optical power variation of $\sim$2~$\mu$W.  
\begin{figure}[tbh!]
\begin{center}
\includegraphics[width=0.7 \columnwidth]{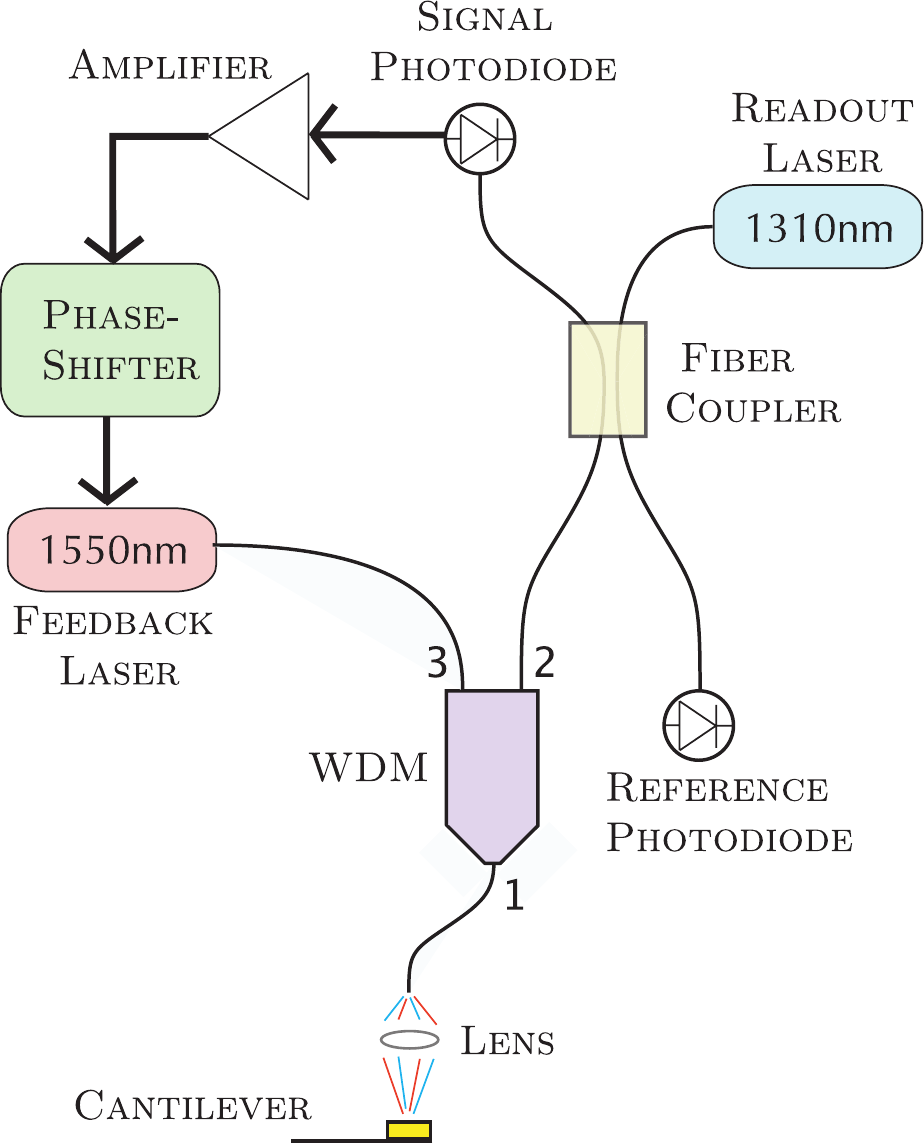}
\end{center}
\vspace{-4mm}
\caption[Radiation pressure damping setup]{(color online). Diagram of the radiation pressure damping system; see text for details.  The wave division multiplexer (WDM) allows 1550 nm light to pass only between ports 1 and 3, and allows 1310 nm light to pass only between ports 1 and 2.} 
\label{setup}
\end{figure}
Our modified fiber-optic interferometer is depicted in Fig.\ \ref{setup}.  The light from the 1310~nm position-detection laser travels through a standard 99/1 fiber coupler and into the ``blue'' arm of a cascaded wave division multiplexer (WDM), then exits the fiber through a flat cleave and is focused by an aspheric lens onto the test mass.  As described above, the reflected power can be converted to a voltage that depends periodically on the cantilever's position.  This voltage is phase-shifted by a custom-built analog circuit and used to modulate the power of a 1550~nm diode laser.  Because photon momentum can only apply force in one direction, it is necessary to add a constant offset to the power so that the force modulation can be of either sign.  The light from the 1550~nm laser is added to the fiber by coupling through the ``red'' arm of the WDM; it then follows the same optical path as the 1310~nm laser, and is focused onto the cantilever by the same optics.  The radiation force exerted on a perfectly reflecting surface by a light beam of power $\mathbf{P}$ is $F_{\mathrm{rad}} = 2\mathbf{P}/c$, where $c$ is the speed of light.  The factor by which $T$ and $Q$ are reduced by feedback is proportional to the gain factor $g$ defined earlier.  The maximum value of $g$ that can be attained using a laser with a maximum rms power modulation amplitude $\langle\mathbf{P}_{\mathrm{mod}}\rangle$ is 
\begin{equation}
g = \frac{2 \langle\mathbf{P}_{\mathrm{mod}}\rangle \omega_{\mathrm{o}}}{c k \langle x \rangle} = \frac{2 \langle\mathbf{P}_{\mathrm{mod}}\rangle \omega_{\mathrm{o}}}{c \sqrt{k k_B T_{\mathrm{o}}}},
\label{dampingequation2}
\end{equation}
where we have used the equipartition theorem to write the root-mean-square cantilever position $\langle x \rangle$ in terms of temperature $T_{\mathrm{o}}$.

\begin{figure}[htb]
\begin{center}
\includegraphics[width=1.0 \columnwidth]{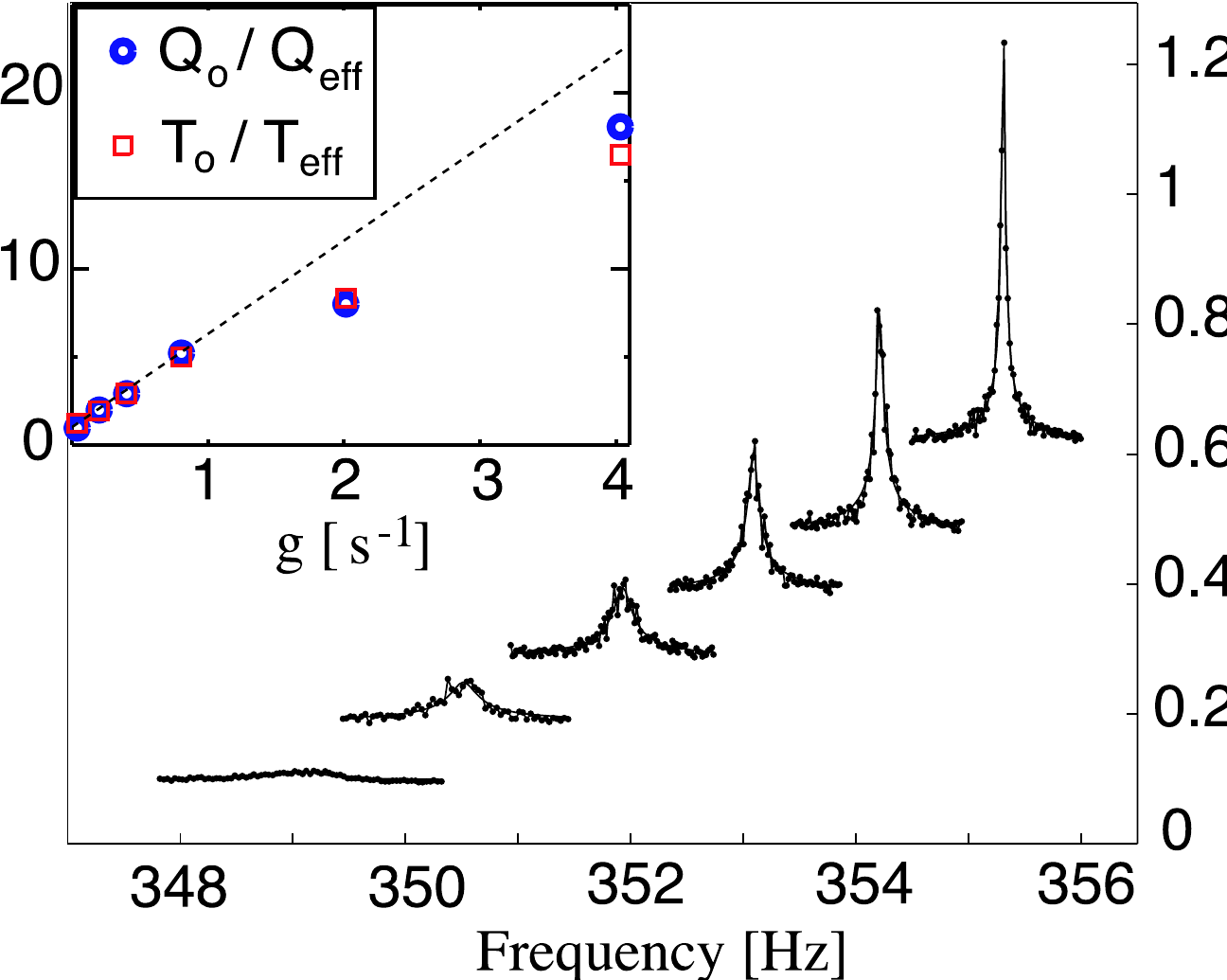}
\end{center}
\vspace{-4mm}
\caption[Radiation pressure damping results]{(color online). Cantilever displacement spectra taken at different feedback gains.  Gain increases from upper right to lower left.  Peaks have been offset in $x$ and $y$ for clarity.  Inset: $Q_{\mathrm{o}}/Q_{\mathrm{eff}}$ (circles) and $T_{\mathrm{o}}/T_{\mathrm{eff}}$ (squares) versus gain factor $g$, for the same data. The dashed line is the theoretical prediction.} 
\label{spectra}
\end{figure}

Results of the feedback-modification of $Q$ and $T$ at room temperature are presented in Fig.\ \ref{spectra}, which shows the broadening and flattening of the thermally excited resonance peak with increasing feedback gain.  The individual spectra were each fitted with a Lorentzian function to extract the value of $Q_{\mathrm{eff}}$.  The effective temperature $T_{\mathrm{eff}}$ was determined by integration of the power spectral density.  Analysis of the lower-leftmost curve shows that the effective temperature of the cantilever was reduced from 300$^\circ$K to 18$^\circ$K, and its quality factor was reduced to $\sim$700.  The measured variation of $Q_{\mathrm{eff}}$ and $T_{\mathrm{eff}}$ with gain is shown in the inset of Fig.\ \ref{spectra}, along with the theoretical prediction.  This reduction of the $Q$ is sufficient to make resonant excitation of a cantilever with the gas bearing rotor possible both at room temperature and at 4.2$^\circ$K.  The ``problem'' of high intrinsic $Q$ discussed above is thus solved without sacrificing force sensitivity.
\section{DATA ACQUISITION AND AVERAGING}
 \label{dataanalysis}
 \subsection{Averaging procedure}
 \label{averaginganalysis} 
There are two main data streams in the experiment: the interferometer waveform and the encoder waveform.  The goal of the averaging procedure is to extract correlations between these two data streams, which represent the test mass position versus time and the drive mass position versus time, respectively.  Both data streams are recorded simultaneously at 10 KHz by the data acquisition system and saved to disk in files that are usually 60 seconds in length.  The data analysis software loads the two main waveforms from each saved data file, breaks them up into smaller segments, and processes each segment to extract position-versus-time curves for the test mass and drive mass.  The drive mass waveform for each segment is subjected to a zero-crossing analysis that determines the rotational frequency of the drive mass, and thus the frequency at which the cantilever would be excited by any real force from the drive mass.  The test mass waveform for each segment is then Fourier transformed, and the amplitude and relative phase of the cantilever's oscillations \emph{at that excited frequency} are determined.  As in a lock-in type measurement, the phase information is crucial, since it enables good noise rejection.   One important parameter of such an analysis is the length of the segments into which the data is divided before the FFT is performed.  We chose, in general, to use segment sizes that corresponded to about three coherence times of the damped cantilever (or about 15 seconds)--- this length was small enough to give us many segments and thus good number statistics, but still large enough so that each segment should be statistically independent.

Any random-phase correlations between the drive mass and test mass positions will be averaged out by this process, and their amplitude will decrease as the inverse root of the averaging time.  Any correlation between the positions that is constant in phase (and is thus not suppressed by the phase-correlated averaging process) presumably represents the influence of a real force coupling the drive mass rotation to the test mass oscillation--- this is the sort of coupling that the experiment is designed to detect or constrain.  
\subsection{Phase determination}		
An important feature of this apparatus is that it has the ability to determine not only the amplitude but also the phase of a measured force between the drive mass and the test mass.  This capability is very useful--- it allows us to distinguish between attractive and repulsive forces, and to identify the presence of certain magnetic forces that appear at other phases. In order to determine the phase of any measured force, the fixed relative phase between the encoder pattern and the mass pattern must be known.  The two patterns must therefore be aligned carefully, and the resulting relative phase must be well-measured.  Alignment of the encoder pattern to the mass pattern is achieved with a shadow mask that is registered to the mass pattern.  Titanium and gold are then deposited through that shadowmask to form the encoder pattern.  After deposition of the encoder pattern but \emph{before} deposition of the opaque gold shield layer visible in Fig.\ \ref{brassmassshield}, the drive mass is photographed so that the phase alignment between the two patterns can be measured and confirmed.  
\begin{figure}[ht!]\begin{center}
\includegraphics[width=0.7 \columnwidth]{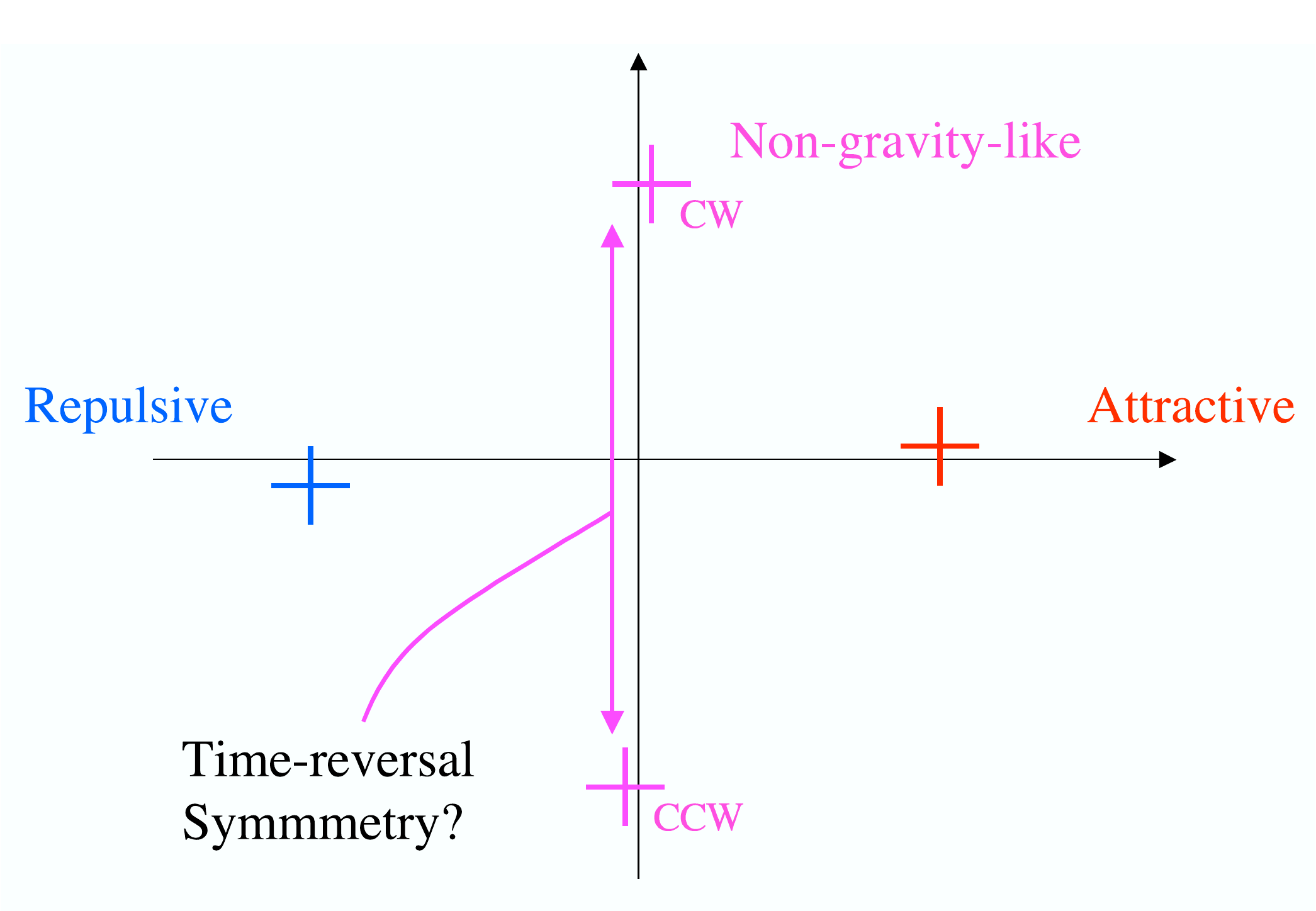}
\caption[Reading data plots.]{(color online). A simple illustration of the meaning of the axes on a typical data plot produced by the analysis described in this section.  The positive real axis corresponds to a purely attractive force.  Time-reversal symmetry breaking will result in different measured forces for CW and CCW runs under otherwise identical experimental conditions.}
\label{readingplots}
\end{center}\end{figure}
\begin{figure}[ht!]\begin{center}
\includegraphics[width=1.0 \columnwidth]{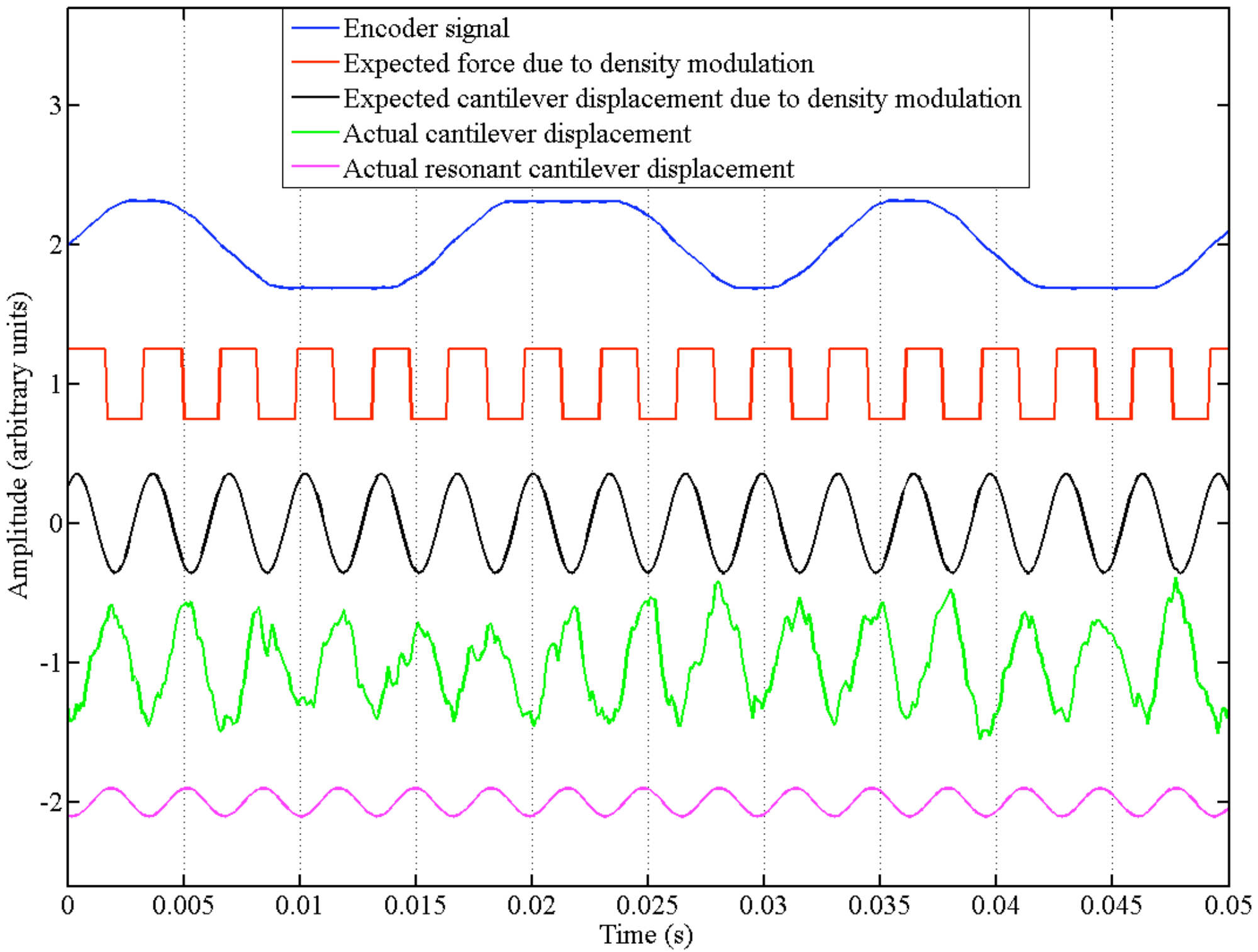}
\caption[Procedure for determining phase of measured force.]{(color online). A plot of the steps used to determine the phase of the measured force, relative to the phase expected from an attractive mass-dependent force.   The top (blue) line is the signal from the optical encoder.  The second (red) line (calculated from the blue line and a knowledge of the relative phase of the two patterns on the drive mass) is the density modulation passing under the cantilever.  The third (black) line is that density modulation put through the complex mechanical transfer function of the cantilever, and thus represents the expected cantilever displacement due to an attractive mass-dependent force.  The fourth (green) line is the measured cantilever displacement.  The fifth (pink) line is the on-resonance Fourier component of 15 seconds of measured cantilever displacement data.  The phase difference between the third line and the fifth line thus represents the phase of the measured force relative to that of an attractive force.  The first four traces are arbitrarily normalized; the last trace is in the same units as the fourth, to show the effect of 15 seconds of averaging.  This graph represents the data from one randomly-chosen 15-second file; results from many such files (with, in general, a wide range of measured phases) are averaged together to provide a meaningful force measurement.}
\label{phasedetermination}
\end{center}\end{figure}
Once the relative phase between the encoder pattern and the mass pattern is known, a series of transformations can be applied to the encoder waveform and the cantilever displacement waveform to determine the phase of the measured force relative to the phase that would be expected from an attractive mass-dependent force like Newtonian gravity. The phase alignment between the mass pattern and encoder pattern is used to derive from the encoder waveform (Fig.\ \ref{phasedetermination}, top trace) the force as a function of time that would be put on the cantilever by an attractive, mass-dependent interaction (Fig.\ \ref{phasedetermination}, second trace).  That force is then Fourier transformed and put through the complex mechanical transfer function of the cantilever, which in frequency space is of the form
\begin{equation}
G(f) = \frac{f_o^2/k}{(f_o^2-f^2)+i(f f_o / Q)},
\end{equation}
where $f$ is frequency, $k$ is the spring constant of the cantilever, $f_o$ is the resonant frequency of the cantilever, $Q$ is the quality factor of the cantilever, and $i$ is $\sqrt{-1}$.  $Q$ and $f_o$ are determined by real-time Lorentzian fitting of the cantilever's resonance peak.  After being put through this function, the force signal is inverse-Fourier-transformed to yield the cantilever displacement versus time that would be expected from a purely attractive force (Fig.\ \ref{phasedetermination}, third trace).  The phase of this signal represents the phase expected from an attractive force. The cantilever displacement measured by the interferometer is then itself Fourier transformed, and the component at the excitation frequency is recorded.  This component has both an amplitude and a phase.  The amplitude represents the amplitude of the measured force at the excitation frequency for this data file.  The phase can be compared to the attractive-force phase that was determined earlier.  This comparison determines whether the measured signal is attractive, repulsive, or somewhere in between.

The end result of this analysis procedure for each segment of data is one point on the complex plane, representing the amplitude and phase (relative to Newtonian gravity) of the measured force.  For a graphical explanation of the meaning of the axes, see Fig.\ \ref{readingplots}.  When many segments of data are taken in a row, they form a distribution on the plane like that visible in Fig.\ \ref{singlerun}.  The mean and standard deviation (divided by the square root of the number of points) of that distribution are then the measured force and the statistical error for that particular run.

\section{NOISE, UNCERTAINTY, AND BACKGROUNDS}

\label{errors}
\subsection{Experimental Uncertainties}
\subsubsection{Uncertainties in mass separation distance}
One of the most important experimental parameters in any Yukawa force measurement is the separation distance between the masses.  In this experiment, that distance is the sum of nine smaller distances, eight of which are fixed by construction.  The ninth is the distance between the top of the drive mass and the bottom of the cantilever wafer, which is variable.  Table~\ref{uncert_massdistance} presents a summary of the estimated uncertainties in each component of the total drive mass to test mass separation.
\begin{table}[htb!]
\begin{center}
\begin{tabular}{lrr}
Distance & Value ($\mu$m) & Error ($\mu$m) \\ \hline \hline
Planarization epoxy thickness & 0.5 & 0.5\\
Drive mass shield layer thickness & 0.4 & 0.1\\
Rotor--wafer separation & 10--20 & 2 \\
Wafer shield coating thickness & 0.4 & 0.1 \\
Wafer shield thickness & 4 & 0.2 \\
Shield--cantilever distance & 14 & 0.1 \\
Cantilever droop & -4 & 1 \\
Cantilever thickness & 0.33 & 0.01 \\
Cantilever--test mass separation & 0 & 1 \\
\hline
Total  & 30.6 & 2.5\\
\hline
\end{tabular}\end{center}
  \caption[Estimated uncertainties in the mass separation]{Table of estimated uncertainties in the determination of the face-to-face separation distance between drive mass and test mass.  The cantilever droop is negative because it decreases the distance between the masses.  The ``total'' numbers assume a 15-$\mu$m rotor-wafer separation. Errors are added in quadrature.}\label{uncert_massdistance}
\end{table}
The distances in Table~\ref{uncert_massdistance} were measured in several different ways.  The  shield thickness, cantilever thickness, and cantilever-shield separation were fixed during fabrication of the cantilever wafer and measured using ellipsometry.  The cantilever-shield separation was also confirmed optically.  The thickness of the two gold layers was measured by a crystal rate monitor during deposition.  The droop of the cantilever in the earth's gravitational field is easily calculated and was confirmed optically.  The cantilever-test mass separation was measured optically.  The thickness of the planarization epoxy layer was calculated \cite{epoxythickness} and confirmed with a profilometer.  Our method for measurement of the variable distance between the drive mass and the cantilever wafer is described in Sec.~\ref{encoder} and depicted in Fig.\ \ref{upanddown}.  
\subsubsection{Uncertainties in the force calibration}
The force on the test mass is inferred from the interferometer voltage using the following relation:
$$
F=\frac{k\ \lambda}{2 \pi\ Q\ \mathrm{V}_\mathrm{pp}}\mathrm{V}_\mathrm{intf},
$$
were $F$ is force, $k$ is the cantilever spring constant, $\lambda$ is the wavelength of the laser, $Q$ is the cantilever's quality factor,  $\mathrm{V}_\mathrm{pp}$ is the interference fringe height (after the preamplifier), and $\mathrm{V}_\mathrm{intf}$ is the measured interferometer voltage.  Table~\ref{uncert_force} presents typical values and uncertainties for these parameters.
\begin{table}[htb!]
\begin{center}\begin{tabular}{lrr}
Quantity & Typical Value & Error  \\ \hline \hline
Quality factor $Q$ (during feedback) & 5000 & 10\%\\ 
Spring constant $k$ (N/m) & 0.04 & 15\%\\
Laser wavelength $\lambda$ (nm)& 1310 & 0.1\%\\
Fringe height $\mathrm{V}_\mathrm{pp}$ (V)& 0.72 & 5\%\\
Interferometer voltage $\mathrm{V}_\mathrm{intf}$ (V) & varies & 2\%\\
\hline
Total error &   & 19\%\\
\hline
\end{tabular}\end{center}
  \caption[Estimated uncertainties in the force calibration]{Table of estimated uncertainties in the voltage-to-force calibration.  Percent errors are added in quadrature.}\label{uncert_force}
\end{table}
\subsubsection{Uncertainties in the mass dimensions and densities}
Any Cavendish-type measurement depends upon knowledge of the density and dimensions of both drive mass and test mass.  Typical values and uncertainties for those parameters are presented in Table~\ref{uncert_massdims}.
\begin{table}[ht!]
\begin{center}\begin{tabular}{lrr}
Quantity & Typical Value & Error  \\ \hline \hline
Test mass density (kg/m$^3$) & 19300 & 3\%\\
Drive mass differential density (kg/m$^3$) & 6200 & 3\%\\
Test mass width ($\mu$m) & 100 & 3\%\\
Test mass length ($\mu$m) & 400 & 3\%\\
Total test mass depth ($\mu$m) & 40 & 15\%\\
Drive mass depth ($\mu$m)  & 1000 & 10\%\\
\hline
\end{tabular}\end{center}
  \caption[Estimated uncertainties in properties of the masses]{Table of estimated uncertainties in important properties of the drive mass and test mass.  Drive mass width and length are not presented because they do not have an important effect on the ultimate determination of the force (the area of overlap between the two masses is determined by the much smaller test mass area). }\label{uncert_massdims}
\end{table}
\subsubsection{Uncertainties in the determination of $\alpha$}
Bounds in  \mbox{$\alpha$-$\lambda$} space on non-Newtonian interactions are derived from the measured force by calculating the value of $\alpha$ that would produce that force at each $\lambda$, given the experimental geometry.  Errors in this procedure can be estimated using an expression for the Yukawa force between two parallel plates, which our experimental geometry closely resembles.  That expression is
$$
F = 2 \pi\ G\ \alpha\ \rho_\mathrm{d}\ \rho_\mathrm{t}\ A\ \lambda^2\ e^{-d/\lambda}\ (1-e^{-t_\mathrm{t}/\lambda})(1-e^{-t_\mathrm{d}/\lambda}),
$$
where $\alpha$ is the strength of the Yukawa force, $\lambda$ is its range, $F$ is the force, $G$ is Newton's constant, $\rho_\mathrm{d}$ and $\rho_\mathrm{t}$ are the densities of the drive mass and test mass respectively, $A$ is the overlapping area of the plates, $d$ is the face-to-face separation between them, and $t_\mathrm{t}$ and $t_\mathrm{d}$ are the thicknesses of the test mass and drive mass, respectively.  Uncertainties in $d$, $F$, and the mass dimensions and densities are discussed above in Tables~\ref{uncert_massdistance}, \ref{uncert_force}, and ~\ref{uncert_massdims}.  Those errors can be propagated through the equation above to calculate the uncertainty in the calculation of $\alpha$ for different values of $\lambda$.  The results are shown in Table~\ref{uncert_alpha}.
\begin{table}[ht!]
\begin{center}\begin{tabular}{rr}
$\lambda$ ($\mu$m) & Error in $\alpha$ (\%)  \\ \hline \hline
1000 & 25.4\\
300 & 24.4\\
100 & 23.5\\
30 & 22.2\\
10 & 28.3\\
3 & 74.4\\
\hline
\end{tabular}\end{center}
  \caption[Estimated uncertainties in calculation of $\alpha$]{Table of estimated uncertainties in the calculation of $\alpha$ for several different values of $\lambda$.}\label{uncert_alpha}
\end{table}

\subsection{Sources of Background Forces}

Because of the extreme weakness of Newtonian gravity, the amplitudes of many of the possible background forces in this experiment are potentially much larger than the signals the experiment is designed to detect.  A successful experiment must then eliminate several different kinds of background forces.  Most non-magnetic forces are reduced either by construction or using the phase-sensitivity of the lock-in-type measurement.  Magnetic forces, which are reduced by the use of non-magnetic materials and mu-metal shielding, will be discussed separately in Sec.~\ref{magforces}.

\subsubsection{Non-magnetic forces}

The main candidates for background forces, Casimir and electrostatic interactions, are eliminated by construction.  As discussed in Sec.~\ref{apparatuschapter}, the two separate layers of gold shielding between test mass and drive mass should entirely eliminate Casimir forces and should greatly reduce electrostatic forces at the period of the mass pattern.  In addition, our use of a phase-sensitive AC ``lock-in'' type measurement offers good protection against backgrounds that are uncorrelated with the density modulation in the drive mass (e.g. from static charges randomly distributed on the drive mass surface).   Corrugation in the drive mass, however, could lead to a spurious force by causing pressure variations that move the shield at a constant phase with respect to the moving mass pattern.  The danger represented by such corrugations is the main reason that the planarization of the drive mass must be done with such precision.  Other gravitational interactions (e.g. the earth's gravity) are completely eliminated by the AC lock-in-type measurement since they are not at all phase-correlated with the drive mass pattern.  The most important possible source of background forces is probably magnetic interactions, which are discussed in detail below.

\subsubsection{Magnetic forces} \label{magforces}
\begin{figure}[ht!]\begin{center}
\includegraphics[width=0.8 \columnwidth]{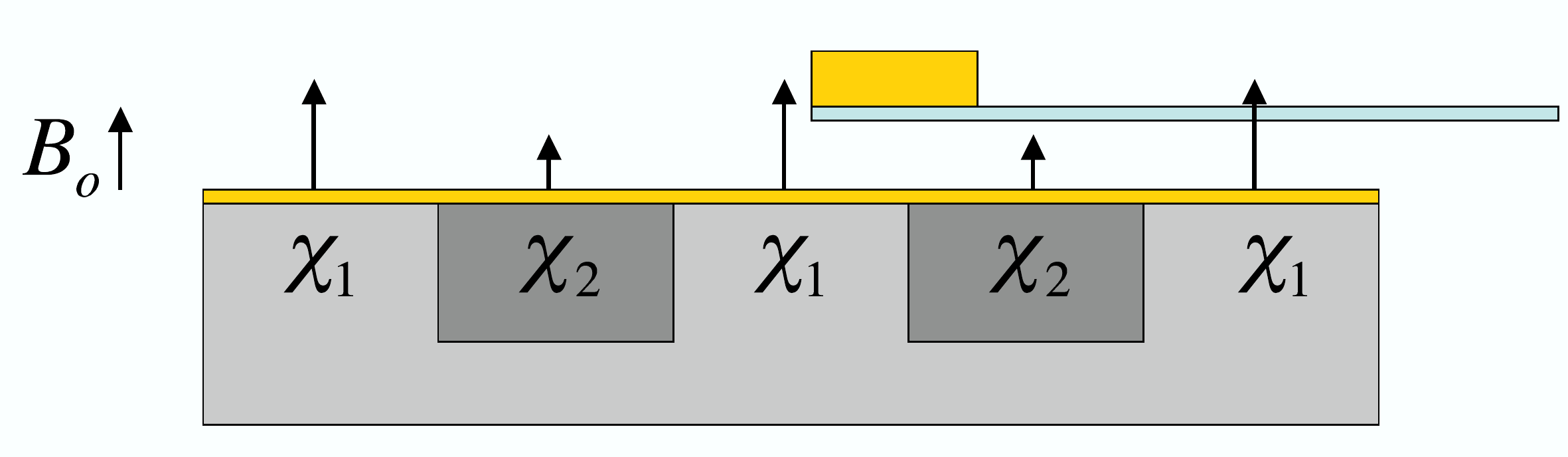}
\caption[Schematic of possible magnetic coupling mechanism]{(color online). A schematic representation of a possible magnetic coupling mechanism that could lead to spurious forces if the drive mass is superconducting.  See text for explanation.}
\label{eddy}
\end{center}\end{figure}
Great care was taken to avoid the use of ferromagnetic materials in the fabrication of the drive and test masses.  However, even in the absence of ferromagnetism, magnetic interactions have the potential to produce systematic errors in a sensitive force measurement.  For example, in the presence of a static magnetic field $B_o$ (possibly due to the earth's field or to trapped flux in superconducting components of the dewar), a drive mass with magnetic susceptibility variation corresponding to the density variation can produce a spurious force.  This force would be large in the case of a superconducting drive mass, since one would then expect a very large susceptibility difference between the dense (superconducting) and light (normal) regions.  See Fig.\ \ref{eddy} for an illustration of the proposed situation.  As the gold test mass passes through the varying magnetic field, it experiences a time-dependent magnetic flux, which causes a non-curl-free electric field due to Faraday's law.  This field causes circulating eddy currents in the test mass.  These eddy currents in turn couple to the magnetic field, resulting in a force on the test mass.  The phase of that force relative to gravity depends to some extent on the details of the coupling, but in general one would expect to see it show up at a phase of $\pm\pi/2$ since it depends on the rate of change of the magnetic field, which should be greatest at the edges of the bars in the drive mass pattern.  

A simple calculation of the expected amplitude of such a force indicates that a force of 10 femtoNewtons could be produced by a field of 100~mG applied to a \emph{superconducting} drive mass.  This assumes perfect diamagnetism in the drive mass, and is thus a conservative upper bound on the possible force.  The actual measured force produced by a superconducting drive mass varies somewhat, and since the magnitude and direction of the force depend on the unknown magnitude and direction of the local magnetic field, no attempt was made to quantitatively investigate the behavior of this force.  Any force of this type will scale with the susceptibility difference between light and heavy regions of the drive mass.  Because of this, a non-superconducting drive mass in the same magnetic field would produce much smaller measured forces.  In particular, when the experiment is operated at a temperature above the critical temperature of lead, the maximum force would thus be reduced by at least four orders of magnitude~\cite{brasssusceptibility}.  All the force-measurement data presented in this paper were taken with the drive mass above its critical temperature.  Further reduction is provided by a mu-metal shield to reduce the ambient magnetic field below 50~mG.  Under normal experimental conditions, then, expected magnetic forces due to induced currents in the test mass are at least three orders of magnitude smaller than the current force resolution of the apparatus, and are not a significant limiting factor.  

\label{trsb}
Interactions that are due to magnetically induced currents, including those discussed above, typically violate time-reversal symmetry.  In the context of this experiment, time-reversal symmetry breaking occurs when the phase and amplitude of the measured force are different for clockwise and counter-clockwise runs. Induced-current forces are not the only ones that break time-reversal symmetry-- viscous forces due to the bearing gas can as well.  Density-dependent forces like Newtonian gravity or Yukawa forces coupling to mass will not break time-reversal symmetry, since the point of maximum force will always be above the dense part of the drive mass pattern.  This suggests a ``Hall-type'' measurement, wherein one measures the force with CW and CCW spin and then adds the results, canceling undesirable time-reversal symmetry breaking terms.  The results of such a measurement are plotted in Fig.\ \ref{dec07runs}.

Magnetic forces other than the type described above, such as forces due to direct coupling of field-induced dipole moments, do not necessarily violate time reversal symmetry.  The amplitude of such forces, though, is also expected to be at least two orders of magnitude smaller than the current force resolution of the apparatus.   Magnetic forces are thus not expected to limit the performance of the experiment.

\subsection{Sources of Noise}

\subsubsection{Thermal noise}

The ultimate limitation on the sensitivity of this experiment is expected to be thermal noise.  Expressed as a lower limit on the detectable force, the thermal noise limit is
$$
F_{\mathrm{detectable}}=\sqrt{\frac{4 k k_\mathrm{B} T b}{Q \omega_o}},
$$
where $k$ is the spring constant of the cantilever, $k_\mathrm{B}$ is Boltzmann's constant, $T$ is the temperature, $b$ is the bandwidth, $Q$ is the cantilever's quality factor, and $\omega_o$ is its frequency.  What this means in practical terms is that a thermal-noise-limited measurement with this apparatus would be (barely) able to resolve Newtonian gravity with a (very achievable) averaging time of 8 hours.  If no anomalous force were detected in such a run, the resulting  \mbox{$\alpha$-$\lambda$} bounds from the measurement would be at least two orders of magnitude stronger than current limits at length scales near 10 $\mu$m.  However, the sensitivity of the experiment is not currently limited by thermal noise, due to the relatively high levels of random vibrational noise.  

\subsubsection{Vibrational noise}

Vibrational noise (random mechanical or acoustic excitation of the cantilever) is currently the factor that limits the force sensitivity of the experiment.  Random vibrational noise increases the scatter of the data points without changing the mean.  There are several possible sources of this type of noise.  The noise we are currently limited by is due to a combination of nitrogen boiloff in the dewar jacket, ambient acoustic noise, and gas flow in the bearing.  The amplitude of this final source depends strongly on bearing flow parameters and temperature, and can be reduced or eliminated by operating in a regime of laminar flow.  The contributions from ambient acoustic noise are minimized with acoustic shielding, and the nitrogen boiloff can be reduced by manipulating the level of liquid helium in the dewar.  The mechanical pump (attached to the dewar through a pump-line isolator) and the mechanical spin of the rotor itself have been shown not to measurably affect the level of vibrational noise.

\subsubsection{Other noise sources}

The experiment in principle contains many sources of noise apart from thermal and vibrational noise, such as Johnson noise in the  $10\ \mathrm{M}\Omega$ feedback resistor of the photodiode amplifier, shot noise in the interferometer and optical damping apparatus, and digitization-induced noise in the data acquisition card.  However, these sources are all expected to put less stringent limits on the sensitivity than thermal noise, and are certainly all much smaller than the current levels of vibrational noise.  Fig.~\ref{spectra} demonstrates the thermal-noise-limited performance of the force-sensing apparatus when it is removed from the main sources of vibrational noise (gas flow and liquid nitrogen boiloff).  Calculated shot noise in the radiation pressure damping laser, for example, is four orders of magnitude smaller than currently measured forces, and could be easily reduced by another factor of 30  by simply decreasing the laser offset power.  Here, we omit to discuss these noise sources in detail since they are not important for the current results and are not expected to limit the sensitivity of the apparatus in the future.

\section{EXPERIMENTAL RESULTS AND ANALYSIS}

\label{results}
 
\begin{figure}[htb!]\begin{center}
\includegraphics[width=0.8 \columnwidth]{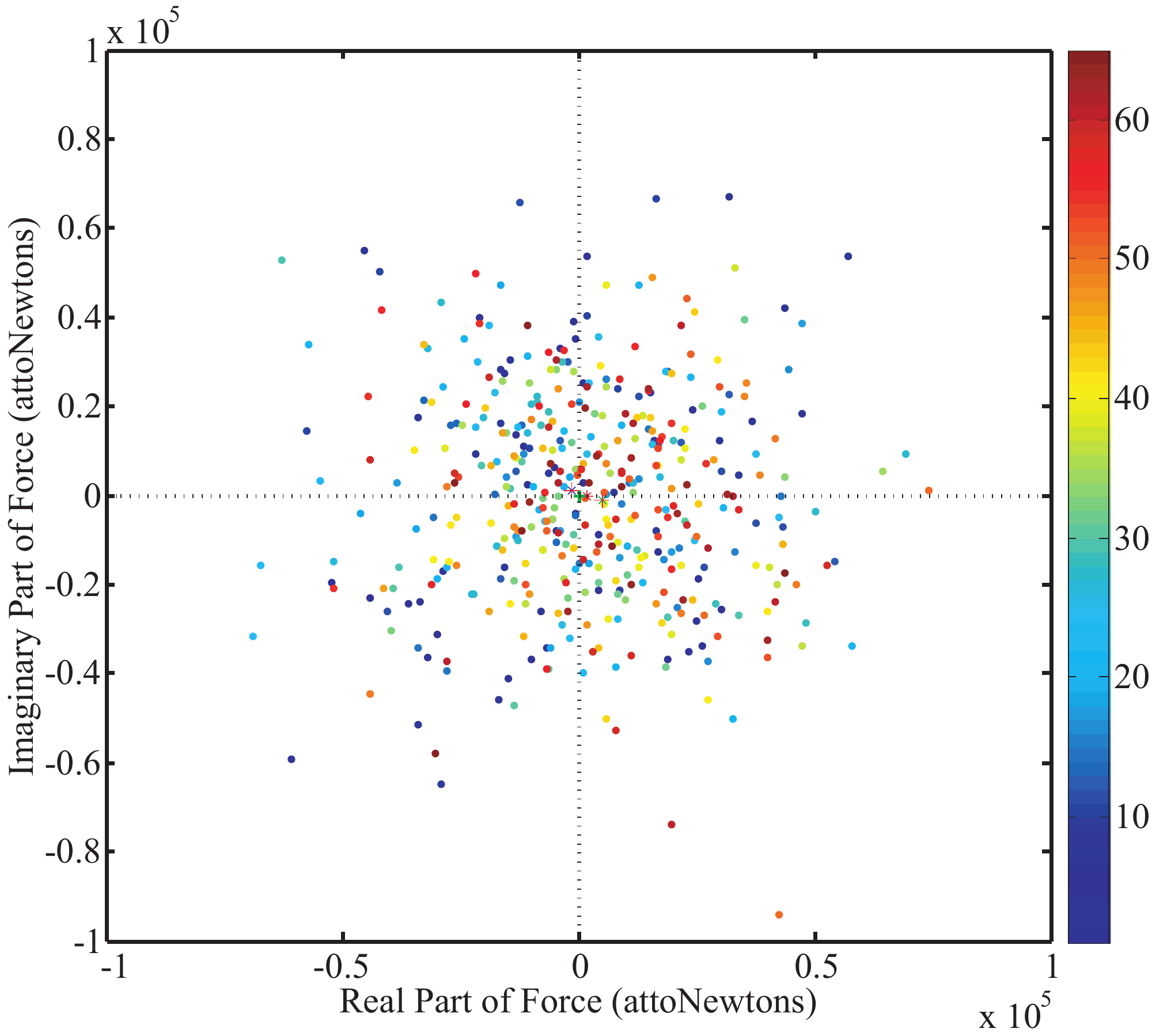}
\caption[Raw data from a single run]{(color online). Mean measured force signals from a single run.  Each point represents 15 seconds of data.  The colors of the points correspond to the order in which they were taken--- higher on the color bar is later in the run.  The mean measured force for this run is consistent with zero at the 2-$\sigma$ level.}
\label{singlerun}
\end{center}\end{figure}

\begin{figure}[htb!]\begin{center}
\includegraphics[width=0.8 \columnwidth]{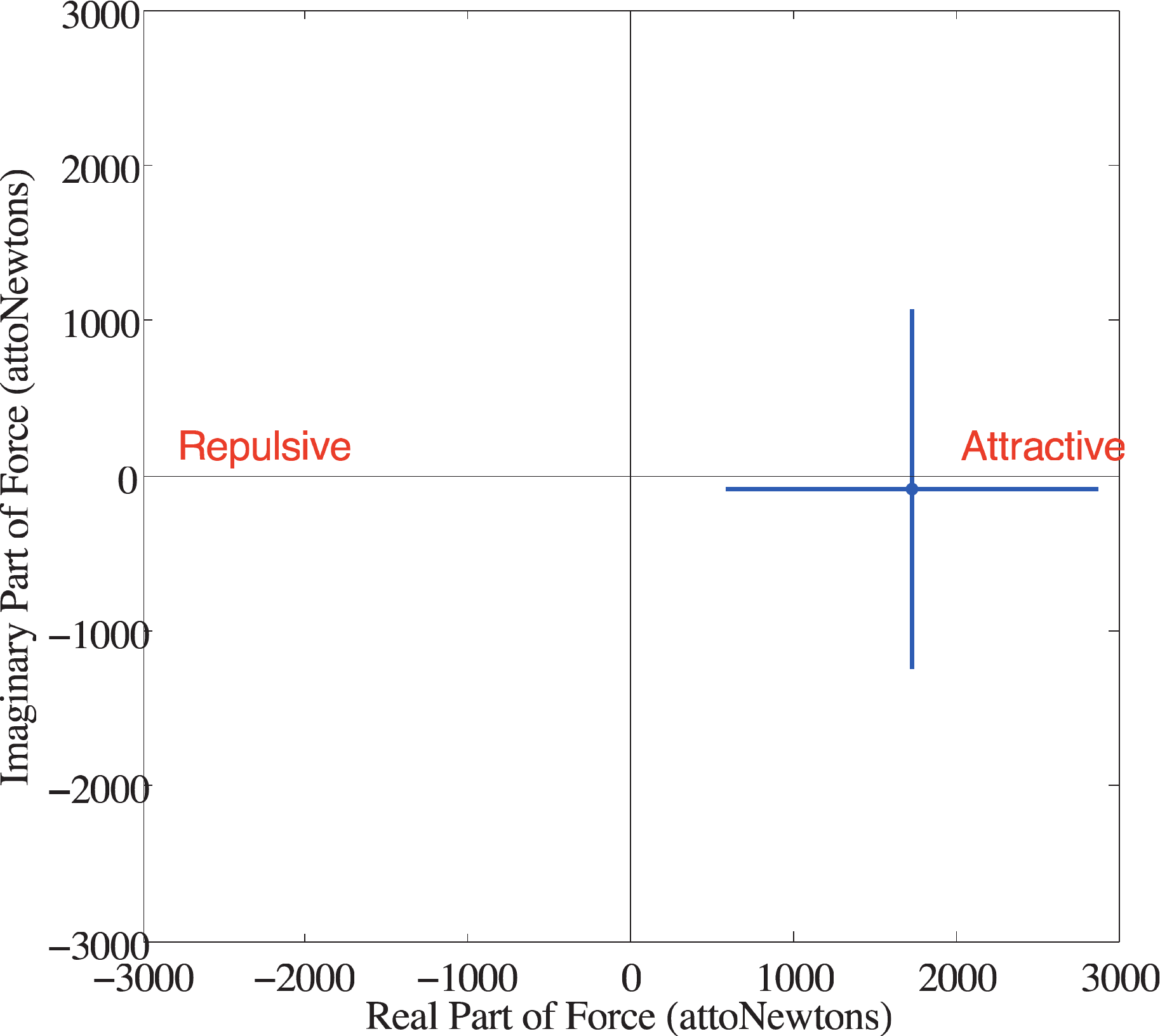}
\caption[Averaged data from a single run]{(color online). Mean and standard deviation of the data plotted in the previous figure.  Error bars are one-$\sigma$ statistical.  The mean measured force for this run is consistent with zero at the 2-$\sigma$ level.}
\label{singlerunaveraged}
\end{center}\end{figure}

This section presents the first data produced by the apparatus.  The data from a single long run lasting 2.5 hours are plotted in Fig.\ ~\ref{singlerun}.  This averaging time was limited only by the need to stop and transfer more liquid helium into the dewar.  The bearing was heated to 8.4$^\circ$K during this clockwise run.  The points are color-coded to indicate the time at which they were taken, and the data were analyzed to determine if there was a drift in the mean force over the course of the run.
Fig.\ \ref{singlerunaveraged} shows the mean force with statistical error bars for the same run as Fig.\ \ref{singlerun}.  The measured force is consistent with zero at the level of two standard deviations (2-$\sigma$).
\begin{figure}[tbh!]\begin{center}
\includegraphics[width=0.8 \columnwidth]{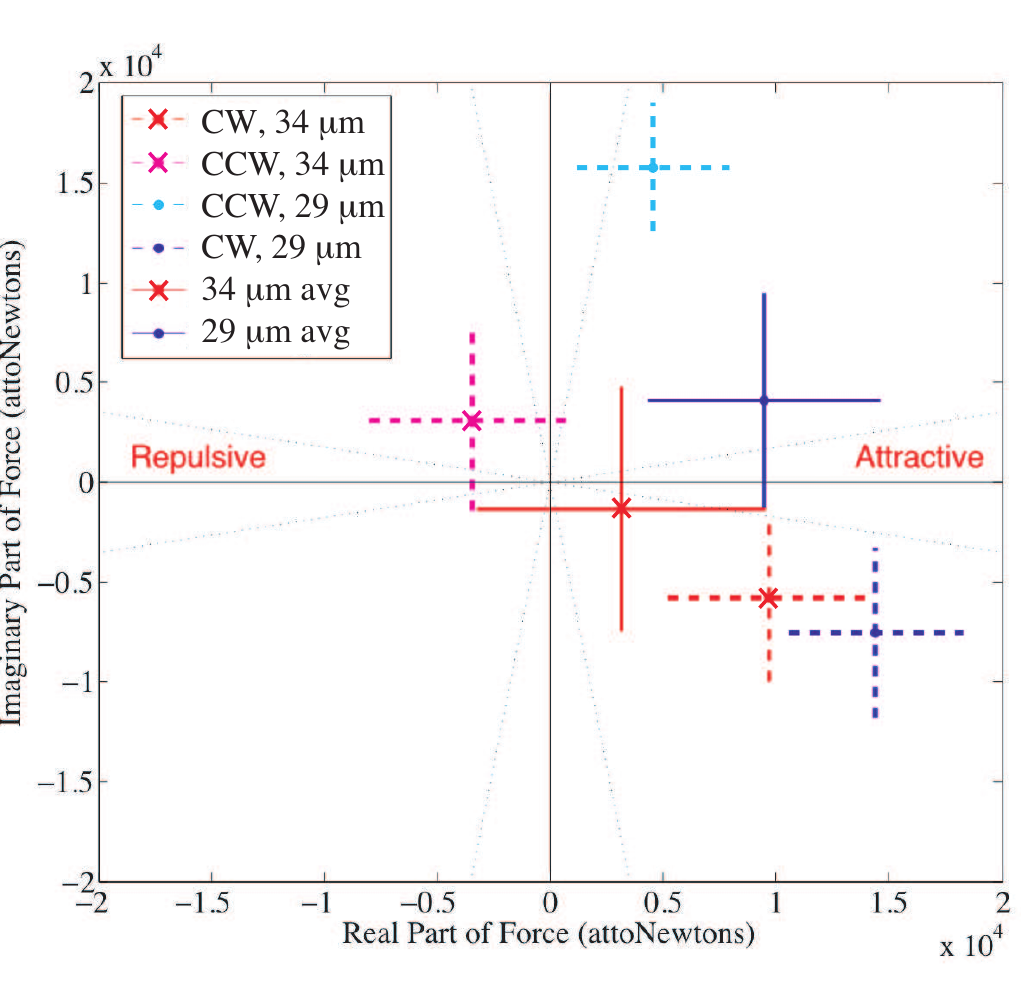}
\caption[Data from multiple runs at different $z$-positions]{(color online). Mean measured force signals from four runs: CW and CCW rotation for each of two different mass separation distances.   The distances in the legend are the measured face-to-face mass separation for each run.  The ``29 $\mu$m'' measurements were taken with 80 sccm flow in the bottom inlet, and the ``34 $\mu$m'' measurements were taken with 56 sccm flow in the bottom inlet. Averages of the two different rotation senses for each separation are also plotted.  Error bars represent one-$\sigma$ statistical errors.}
\label{dec07runs}
\end{center}\end{figure}
In order to help elucidate the effects of different bearing flow parameters and further test the time-reversal symmetry of the measurement, four sets of data were taken in one day.  These data were taken with the drive mass rotating both clockwise and counter-clockwise for each of two different values of flow in the bottom gas inlet (56 and 80 sccm He).  The flow in the top gas inlet was held at 40 sccm He for all four sets, and the bearing exhaust temperature was held at 9$^\circ$K.  The two different flow settings resulted in two different equilibrium $z$-positions for the rotor: 18 $\mu$m and 13 $\mu$m below the top, respectively.  These positions corresponded to face-to-face mass separations of 34 $\mu$m and 29 $\mu$m (see Fig.\ \ref{upanddown} for an explanation of our method for changing the mass separation distance).  The results of these runs are shown in Fig.\ \ref{dec07runs}.  As before, only the mean force for each run is plotted, along with one-$\sigma$ statistical error bars.  Also shown are the average of the CW and CCW runs for each value of the bottom gas flow.  This sort of averaging should cancel out effects which violate time-reversal symmetry.

The data show a clear effect from reversing the sense of rotation.  The imaginary part of the measured force (corresponding to a phase 90$^\circ$ from that of Newtonian gravity) switches sign between the CW and CCW runs.  In fact, for the 34-$\mu$m-separation runs, the CW and CCW measurements are symmetric around the origin, within the one-$\sigma$ statistical error.  The mean force of the two 29-$\mu$m-separation runs is slightly displaced in the direction of attractive force, although it is also consistent with zero, this time at the \emph{two}-$\sigma$ level.  The statistical errors for these runs are large because they did not last very long; each of the four runs represents only about 45 minutes of data.  The short length was due to the time constraints imposed by the need to take four runs on one dewar of helium.  These data give a blueprint for further exploration of experimental phase space.
\subsection{General properties of data}
Although the experimental parameters were varied fairly widely in the name of optimization, the data observed so far show a few generally invariant properties.  The most obvious of these is the preponderance of positive real parts--- every data run that had a well-defined phase had a real part greater than zero.  Thus, the measured force, when phase-correlated with the mass pattern, was always attractive.  This effect  was not always statistically significant, however.

The other property that nearly all the experimental runs have shared is some level of time-reversal symmetry breaking.  The imaginary component of the measured force tended to be positive for counter-clockwise runs and negative for clockwise runs. 
\subsection{Limits on non-Newtonian interactions}
\begin{figure}[tbh!]\begin{center}
\includegraphics[width=1.0 \columnwidth]{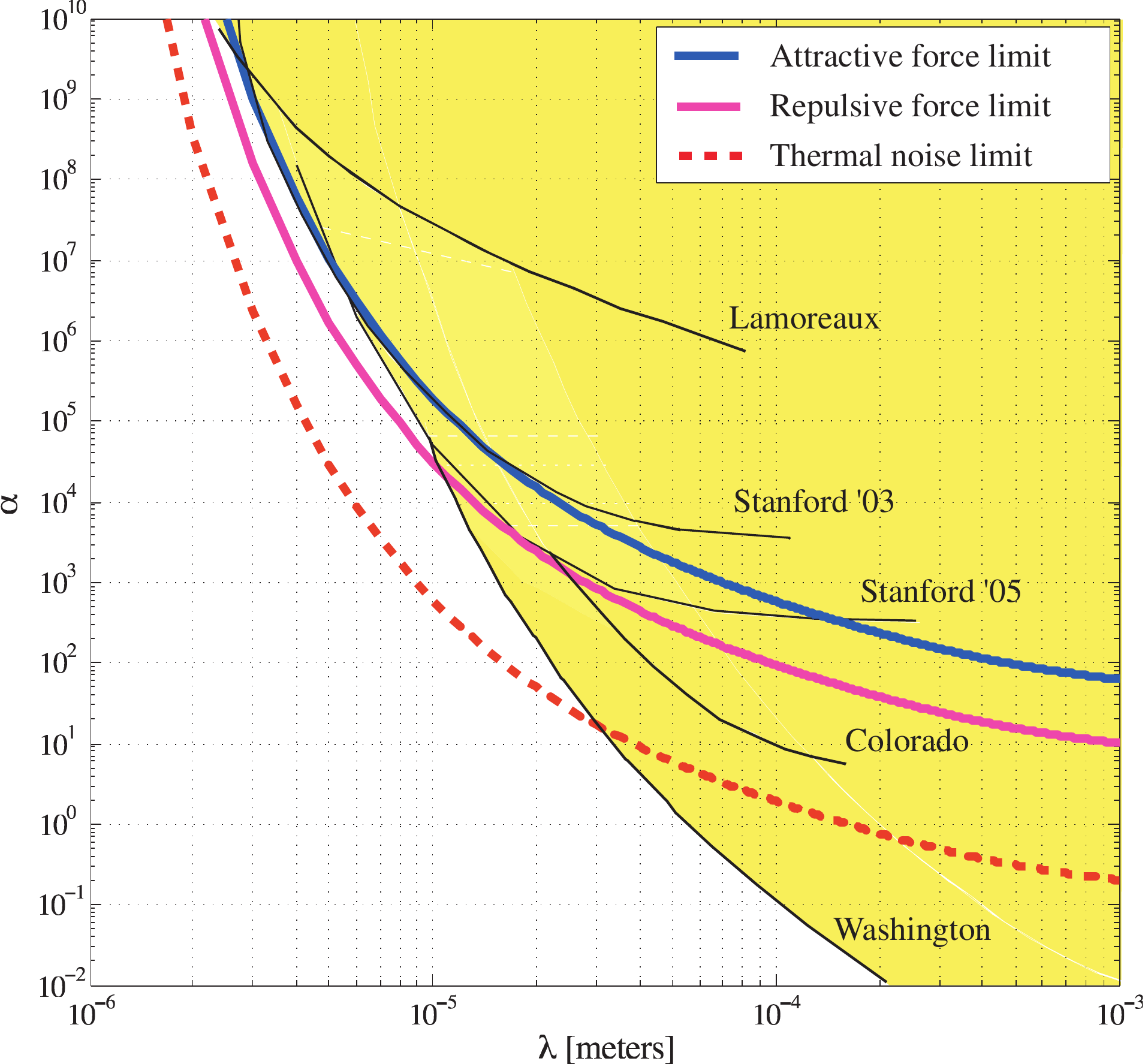}
\caption[Limits in $\alpha--\lambda$ space]{(color online). The bounds on non-Newtonian interactions extracted from the first data taken with this experiment.  Bounds on attractive interactions are represented by the solid blue line, bounds on repulsive interactions by the solid pink line.  Each line represents the mean measured force with experimental uncertainties and a two-$\sigma$ statistical error.  The data from Refs.\ \cite{frogland1,frogland2,adelbergerexpt1,priceexpt,lamoreauxalphlam,newadelberger} are plotted for reference.  The bounds that would be achieved by thermal-noise-limited operation of the present configuration of this apparatus are shown as a dashed line.}
\label{resultalphlam}
\end{center}\end{figure}
The data plotted in Fig.\ \ref{singlerunaveraged} can be combined with data from similar runs (for a total of five hours of integration time) to provide bounds on non-Newtonian interactions.  The bounds are extracted by calculating the expected force between the masses that would result from a Yukawa interaction with a particular $\alpha$ and $\lambda$.  For each $\lambda$, the measured upper bound on $\alpha$ is the value of $\alpha$ at which that Yukawa force would be excluded by our measurements at the 2-$\sigma$ level.  The resulting bounds are plotted in  \mbox{$\alpha$-$\lambda$} space in Fig.\ \ref{resultalphlam}.  Limits on both attractive and repulsive forces have been plotted.  The presence of two bounds is a departure from tradition; typically the $\alpha$ on  \mbox{$\alpha$-$\lambda$} plots represents the absolute value of $\alpha$.  The point of plotting the pink line is to show that, since the mean measured force is attractive, the range of negative $\alpha$ ruled out at the 2-$\sigma$ level is greater than that of positive $\alpha$'s.  The limits on attractive forces that can be extracted from the first runs of this experiment are equal to the best existing limits at five-micron length scales.  The limits on repulsive forces, as mentioned above, are better.  As shown in Fig.\ \ref{resultalphlam}, the ultimate thermal-noise-limited sensitivity of this experiment would be substantially better than any limit achieved so far at these scales.  

\section{CONCLUSIONS \& FUTURE PROSPECTS}

We have designed, constructed, tested, and run a new experiment for detecting or constraining deviations from Newtonian gravity at small length scales.  To overcome some of the challenges associated with measuring gravitational interactions at short distances, we have developed several techniques that could prove to be useful in similar experiments or other areas.  In particular, the method of using radiation pressure to cool a micromachined cantilever, the combination of encapsulated MEMS sensors with an integrated cryopump, the use of a macroscopic gas bearing as a general-purpose low-temperature motor, the methods of simple and precise fabrication of intermediate-scale metallic structures, and some of the techniques of dimensionally stable planarization could find wider application.  

The first experimental runs of the apparatus produced the limits on non-Newtonian interactions shown in Fig.\ \ref{resultalphlam}.  The limits on attractive forces thus obtained are equal to the best limits at 5-micron length scales, and the limits on repulsive forces are an improvement on existing bounds by a factor of 5.  The force sensitivity is currently limited by vibrational noise.   Fig.\ \ref{resultalphlam} also shows the excellent potential thermal-noise-limited sensitivity of this apparatus.  

We thank Dan Rugar for useful discussions, Savas Dimopoulos for theoretical perspectives, and Sylvia Smullin for a critical reading of the manuscript.  This work was supported by NSF grant number PHY-0554170.


\end{document}